\documentclass[prb,twocolumn,showpacs,preprintnumbers,amsmath,aps]{revtex4}
\usepackage{graphicx,hyperref}
\usepackage{bm}
\usepackage{subfigure}

\newcommand\ie{{\it i.e.}}

\newcommand\vect[1]{ \boldsymbol{ #1}}
\def\nbN{\ensuremath{\mathrm{I\!N}}} 
\def\nbC{{\mathchoice {\setbox0=\hbox{$\displaystyle\rm C$}%
\hbox{\hbox to0pt{\kern0.4\wd0\vrule height0.9\ht0\hss}\box0}}
{\setbox0=\hbox{$\textstyle\rm C$}\hbox{\hbox
to0pt{\kern0.4\wd0\vrule height0.9\ht0\hss}\box0}}
{\setbox0=\hbox{$\scriptstyle\rm C$}\hbox{\hbox
to0pt{\kern0.4\wd0\vrule height0.9\ht0\hss}\box0}}
{\setbox0=\hbox{$\scriptscriptstyle\rm C$}\hbox{\hbox
to0pt{\kern0.4\wd0\vrule height0.9\ht0\hss}\box0}}}}
\def\nbQ{{\mathchoice {\setbox0=\hbox{$\displaystyle\rm
Q$}\hbox{\raise
0.15\ht0\hbox to0pt{\kern0.4\wd0\vrule height0.8\ht0\hss}\box0}}
{\setbox0=\hbox{$\textstyle\rm Q$}\hbox{\raise
0.15\ht0\hbox to0pt{\kern0.4\wd0\vrule height0.8\ht0\hss}\box0}}
{\setbox0=\hbox{$\scriptstyle\rm Q$}\hbox{\raise
0.15\ht0\hbox to0pt{\kern0.4\wd0\vrule height0.7\ht0\hss}\box0}}
{\setbox0=\hbox{$\scriptscriptstyle\rm Q$}\hbox{\raise
0.15\ht0\hbox to0pt{\kern0.4\wd0\vrule height0.7\ht0\hss}\box0}}}}
\def\nbT{{\mathchoice {\setbox0=\hbox{$\displaystyle\rm
T$}\hbox{\hbox to0pt{\kern0.3\wd0\vrule height0.9\ht0\hss}\box0}}
{\setbox0=\hbox{$\textstyle\rm T$}\hbox{\hbox
to0pt{\kern0.3\wd0\vrule height0.9\ht0\hss}\box0}}
{\setbox0=\hbox{$\scriptstyle\rm T$}\hbox{\hbox
to0pt{\kern0.3\wd0\vrule height0.9\ht0\hss}\box0}}
{\setbox0=\hbox{$\scriptscriptstyle\rm T$}\hbox{\hbox
to0pt{\kern0.3\wd0\vrule height0.9\ht0\hss}\box0}}}}
\def\nbS{{\mathchoice
{\setbox0=\hbox{$\displaystyle     \rm S$}\hbox{\raise0.5\ht0%
\hbox to0pt{\kern0.35\wd0\vrule height0.45\ht0\hss}\hbox
to0pt{\kern0.55\wd0\vrule height0.5\ht0\hss}\box0}}
{\setbox0=\hbox{$\textstyle        \rm S$}\hbox{\raise0.5\ht0%
\hbox to0pt{\kern0.35\wd0\vrule height0.45\ht0\hss}\hbox
to0pt{\kern0.55\wd0\vrule height0.5\ht0\hss}\box0}}
{\setbox0=\hbox{$\scriptstyle      \rm S$}\hbox{\raise0.5\ht0%
\hboxto0pt{\kern0.35\wd0\vrule height0.45\ht0\hss}\raise0.05\ht0%
\hbox to0pt{\kern0.5\wd0\vrule height0.45\ht0\hss}\box0}}
{\setbox0=\hbox{$\scriptscriptstyle\rm S$}\hbox{\raise0.5\ht0%
\hboxto0pt{\kern0.4\wd0\vrule height0.45\ht0\hss}\raise0.05\ht0%
\hbox to0pt{\kern0.55\wd0\vrule height0.45\ht0\hss}\box0}}}}
\def\nbZ{{\mathchoice {\hbox{$\sf\textstyle Z\kern-0.4em Z$}}
{\hbox{$\sf\textstyle Z\kern-0.4em Z$}}
{\hbox{$\sf\scriptstyle Z\kern-0.3em Z$}}
{\hbox{$\sf\scriptscriptstyle Z\kern-0.2em Z$}}}}

\begin{document}

\title{Non-Perturbative Functional Renormalization Group for Random
  Field Models and Related Disordered Systems. II: Results for the
  Random Field $O(N)$ Model}

\author{Matthieu Tissier} \email{tissier@lptl.jussieu.fr}
\altaffiliation{Present address: Instituto de F\'\i{}sica, Facultad de
ingener\'\i{}a, Universidad de la Rep\'ublica, J.H. y Reissig 565,
11000 Montevideo, Uruguay}
\affiliation{LPTMC, CNRS-UMR 7600, Universit\'e Pierre et Marie Curie,
bo\^ite 121, 4 Pl. Jussieu, 75252 Paris c\'edex 05, France}

\author{Gilles Tarjus} \email{tarjus@lptl.jussieu.fr}
\affiliation{LPTMC, CNRS-UMR 7600, Universit\'e Pierre et Marie Curie,
bo\^ite 121, 4 Pl. Jussieu, 75252 Paris c\'edex 05, France}

\date{\today}

\begin{abstract}
  We study the critical behavior and phase diagram of the
  $d$-dimensional random field $O(N)$ model by means of the
  nonperturbative functional renormalization group approach presented
  in the preceding paper. We show that the dimensional reduction
  predictions, obtained from conventional perturbation theory, break
  down below a critical dimension $d_{DR}(N)$ and we provide a
  description of criticality, ferromagnetic ordering and quasi-long
  range order in the whole $(N,d)$ plane. Below $d_{DR}(N)$, our
  formalism gives access to both the typical behavior of the system,
  controlled by zero-temperature fixed points with a nonanalytic
  dimensionless effective action, and to the physics of rare
  low-energy excitations (``droplets''), described at nonzero
  temperature by the rounding of the nonanalyticity in a thermal
  boundary layer.
\end{abstract}

\pacs{11.10.Hi, 75.40.Cx}

\maketitle

\section{Introduction}
\label{sec:introduction}

The random field model describes one of the simplest disordered
systems in which classical $N$-component variables (spins in magnetic
language) with $O(N)$ symmetric interactions are linearly coupled to a
random (magnetic) field. Yet, more than thirty years after the first
studies of the model,\cite{imry75,aharony76,grinstein76} its long-distance
behavior (criticality and ordering) still largely appears as a puzzle.

The two main questions raised about the (equilibrium) properties
concern the nature and the characteristics of the phases and of the
phase transitions. The first one is about the so-called ``dimensional
reduction'' property which relates the critical behavior of the random
field $O(N)$ model (RF$O(N)$M) in dimension $d$ to that of the pure
$O(N)$ model in dimension $d-2$. This property, predicted to all
orders by conventional perturbation
theory\cite{grinstein76,aharony76,young77} and derived as a
consequence of a hidden supersymmetry,\cite{parisi79} is known to
break down in low enough dimensions.\cite{imbrie84,bricmont87} The
second question concerns the existence of a phase with quasi-long
range order (QLRO) (\ie, a phase characterized by no magnetization and
a power-law decrease of the correlation fuctions) in the models with a
continuous symmetry ($N>1$) below $d=4$, their lower critical
dimension for long-range ferromagnetism. More specifically, the
presence of QLRO in the $3$-dimensional RF$XY$M ($N=2$) is of
relevance to the ``Bragg glass'' phase discussed in the context of
vortices in disordered type-$II$
superconductors.\cite{giamarchi94,blatter94,giamarchi95,giamarchi98,nattermann00}

In the preceding article,\cite{tarjus07_1} denoted as paper I in the
following, we have developed a \textit{nonperturbative functional
  renormalization group} (NP-FRG) formalism to study the long-distance
physics of random field models and related disordered systems. In the
present article, we put the formalism to use to address the issues
mentioned above concerning the behavior of the RF$O(N)$M.

We start in section II by briefly recalling the main definitions,
notations and results of the NP-FRG approach to the RF$O(N)$M which
have been presented in paper I.

Next, in section III, we discuss the mechanism by which dimensional
reduction breaks down, namely the appearance of a strong enough
nonanalytic behavior in the field dependence of the dimensionless
effective average action. We first recall the analysis of the
perturbative functional RG at one loop near $d=4$. We then present the
scenario for the failure of dimensional reduction in the RFIM within
our NP-FRG approach and extend our considerations to the whole $(N,d)$
plane.

The numerical results obtained from our minimal trucation of the
NP-FRG for the RF$O(N)$M are presented in section IV. This allows us
to provide a unified description of criticality, ferromagnetism, and
QLRO in the whole $N-d$ diagram. We show that two nontrivial critical
lines characterize the long-distance behavior of the RF$O(N)$M (on top
of the upper, $d_{uc}=6$, and lower, $d_{lc}(N=1)=2$ and
$d_{lc}(N>1)=4$, critical dimensions for the paramagnetic to
ferromagnetic transition): a line $d_{DR}(N)$ separating a region of
the $(N,d)$ plane in which the critical exponents are given by the
dimensional reduction predictions ($d>d_{DR}(N)$) from a region where
dimensional reduction is fully broken ($d<d_{DR}(N)$); and a line
$d_{lc}(N)$ characterizing the lower critical dimension for quasi-long
range order (for $1<N\leq N_{c}=2.83...$). Finally, we discuss the
accuracy and reliability of the present truncation.

In section V, we address the physical meaning of the nonanalyticity
(in the effective action) which is associated with dimensional
reduction failure. We also examine the role of temperature and the
connection with the phenomenological droplet approach. In particular,
we discuss the ``activated dynamic scaling'' behavior that
characterizes the critical slowing down of relaxation in the RFIM.

We finally conclude by considering the relation of the present to
other pictures of the behavior of random field models and providing
some perspectives.

Short accounts of the present work have already been published in
Refs.~[\onlinecite{tarjus04,tissier06}].

\section{NP-FRG approach for the RF$O(N)$M}
\label{sec_npfrg}

The NP-FRG approach developed in paper I combines three main ingredients:
\begin{enumerate}
\item a version of Wilson's continous RG in which one follows the
  evolution of the effective average action $\Gamma_k$ with a
  (momentum) scale $k$, from the bare action at the microscopic scale
  ($k=\Lambda$) to the full effective action at macroscopic scale
  ($k=0$); the evolution of $\Gamma_k$, which is the generating
  functional of the $1$-particle irreducible vertices at scale $k$, is
  governed by an exact RG flow equation.\cite{berges02}

\item A replica formalism in which the permutational symmetry among
  replicas is explicitly broken by the introduction of linear sources
  acting independently on each replica; using an expansion in the
  number of unconstrained (or ``free'') replica sums gives access to a
  description of the probability distribution of the renormalized
  disorder through its cumulants.

\item A nonperturbative approximation scheme for the effective average
  action that relies on truncating both it ``derivative expansion''
  (expansion in the number of spatial derivatives of the fundamental
  fields) and the ``expansion in number of free replica sums'' (or,
  equivalently, the cumulant expansion).
 
\end{enumerate}

For the RF$O(N)$M, the minimal truncation of $\Gamma_k$ which already
contains the key features for a nonperturbative study of the
long-distance physics is the following:
\begin{equation}
\label{eq_Gamma_k}
\begin{split}
  \Gamma_k\left[\{\vect \phi_a\}\right ]= \int_{\vect x} \bigg\{ \frac
  1{2} \sum_{a=1}^n &Z_{m,k} \vert \partial \vect \phi_a(\vect x)\vert
  ^2 + \sum_{a=1}^n U_{k}(\vect \phi_a(\vect x)) \\&-\frac 1{2}
  \sum_{a,b=1}^n V_{k}(\vect \phi_a(\vect x), \vect \phi_b(\vect x))
  \bigg\},
\end{split}
\end{equation}
where as before $\vect \phi_a$, $a=1,...,n$, are the replica fields,
$Z_{m,k}$ is a wave function renormalization parameter, $U_k$ is the
$1$-replica potential which physically represents a coarse-grained
Gibbs free energy and gives access to the thermodynamics of the system,
and $V_k$ is the $2$-replica potential which is the second cumulant of
the renormalized disorder evaluated for uniform fields.

The flow equations for $U_k(\vect \phi_1)$, $V_{k}(\vect \phi_1, \vect
\phi_2)$, and $Z_{m,k}$ are obtained from the exact RG equation for
the effective average action,\cite{berges02,tarjus07_1}
\begin{equation}
\label{eq_erg}
\partial_k\Gamma_k\left[\{\vect \phi_a\}\right ]=
\dfrac{1}{2} \int_{\vect q} Tr \left\lbrace \partial_k \vect R_k(q^2) \left[\vect \Gamma _k^{(2)}+\vect R_k\right]_{\vect q,- \vect q}^{-1}\right\rbrace ,
\end{equation}
where the trace involves a sum over both replica indices and
$N$-vector components and $\vect \Gamma_k^{(2)}$ is the tensor formed
by the second functional derivatives of $\Gamma_k$ with respect to the
fields $\phi_a^\mu(\vect q)$. $\vect R_k(q^2)$ is the infrared cutoff
which enforces the decoupling of the low- and high-momentum modes at
the scale $k$. It is diagonal in $N$-vector indices, and in the
minimal truncation we have also chosen it diagonal in replica indices,
\textit{i.e.}, $R_{k,ab}^{\mu\nu}(q^2) = \widehat{R}_k(q^2)\delta_{ab}
\delta_{\mu\nu}$.

The initial condition is given by the bare replicated action of the RF$O(N)$M,
\begin{equation}
\begin{split}
  S\left[\{\vect \phi_a\}\right ]&= \int_{\vect x}\bigg\{\frac
  {1}{2T}\sum_{a=1}^n\big[\vert \partial \vect \phi_a(\vect x)\vert
  ^2+\tau \vert \vect \phi_a(\vect x) \vert^2+\\& \frac u{12}(\vert
  \vect \phi_a(\vect x)\vert^2) ^2\big]-\frac {\Delta}{2T^2}
  \sum_{a,b=1}^n \vect \phi_a(\vect x)\cdot \vect \phi_b(\vect x)
  \bigg\},
\end{split}
\end{equation}
where we have made explicit the dependence on a ``bare'' temperature $T$.

One more step is needed to cast the NP-FRG flow equations in a form
suitable for searching for the anticipated zero-temperature fixed
points of the RF$O(N)$M,\cite{villain84,fisher86b} namely, to introduce appropriate scaling
dimensions. This requires to define a renormalized temperature $T_k$
which is expected to flow to zero as $k \rightarrow 0$. Near a
zero-temperature fixed point, one has the following scaling
dimensions:
\begin{equation}
T_k \sim k^\theta,\;  Z_{m,k} \sim k^{-\eta}, \; \phi_a^\mu \sim k^{\frac{1}{2}(d-4+\bar \eta)},
\end{equation}
with $\theta$ and $\bar \eta$ related through $\theta=2+\eta-\bar \eta$, as well as
\begin{equation}
U_k\sim k^{d-\theta}, \;V_k \sim k^{d-2\theta},
\end{equation}
so that the second cumulant of the renormalized random field,
\begin{equation}
\Delta_k^{\mu\nu}(\vect \phi_1,\vect \phi_2)= \partial \phi_1^\mu \partial \phi_2^\nu V_k(\vect \phi_1,\vect \phi_2),
\end{equation}
scales as $k^{-(2\eta- \bar \eta)}$. (We recall that the superscripts with greek letters denote the components of the $N$-vector  fields.)

The dimensionless counterparts of $U_k, V_k,\Delta _k, \vect \phi$ are
denoted by lower-case letters, $u_k, v_k,\delta _k, \vect
\varphi$. (For the RF$O(N)$M, a convenient parametrization of the $1$-
and $2$-replica functions makes use of the variables $\rho=\vert \vect
\varphi \vert^2/2$ and $z=\vect \varphi_1 \cdot\vect \varphi_2
/\sqrt{4\rho_1\rho_2}$.) The resulting flow equations in scaled form
have been given in section IV of paper I.

We conclude this brief recapitulation of the results of paper I by
recalling that the minimal nonperturbative truncation described above
reduces to the $1$-loop perturbative results near the upper critical
dimension $d=6$ and when $N \rightarrow \infty$, and, most
importantly, that it reproduces the $1$-loop perturbative FRG
equations near the lower critical dimension for ferromagnetism in the
$O(N>1)$ model, $d=4$.

\section{Breakdown of dimensional reduction}
\label{sec_dimensional reduction}

As stressed in the introduction, dimensional reduction for the random
field model must break down in low enough dimension. Within the
NP-FRG, we find that the mechanism by which this occurs is the
appearance along the RG flow of a nonanalyticity in the field
dependence of the dimensionless effective average action; more
precisely, the appearance of a cusp in the second cumulant of the
renormalized random field as one makes the two field arguments
approach each other. The theory is renormalizable, albeit with the
unusual feature that the renormalized effective action is nonanalytic
at the fixed point. Such a mechanism has previously been found in the
random manifold model within the perturbative FRG approach.\cite{fisher86,balents93,balents96,ledoussal02b,ledoussal04}

In this section, we discuss the appearance of a nonanalytic behavior
within our minimal truncation scheme. We first recall the results
obtained near $d=4$ for the
RF$O(N>1)$M,\cite{fisher85,feldman02,tarjus04,tissier06b} since this
limit is more easily accessible to an analytic treatment and already
provides the scenario for the general case.

\subsection{RF$O(N)$M at one loop near $D=4$}

Our starting point is the set of equations derived at first order in
$\epsilon=d-4$ and at zero temperature for the running exponents
$\eta_k$, $\bar \eta_k$, and for the dimensionless renormalized second
cumulant of the disorder, more precisely for $R_k(z) =
v_k(\rho_{m,k},\rho_{m,k},z)/(2\rho_{m,k})^2$ where $\rho_{m,k}$
corresponds to the minimum of the $1$-replica potential
(\textit{i.e.}, is akin to a dimensionless order parameter at the
running scale $k$) and goes as $1/\epsilon$ at the relevant fixed
points: see Eqs. (100) and (101) of paper I. (Note that $R_k(z)$
should not be confused with the regulator $\textbf{R}_k(q^2)$ that
appears for instance in Eq.~(\ref{eq_erg}).)  For studying the fixed
points and their stability it is convenient to introduce
$\widetilde{R}_k(z)=(4v_4/\epsilon)R_k(z)$ and to rescale the RG
``time'' as $\epsilon t \rightarrow t$. The flow equation then reads
\begin{equation}
  \label{eq_tildeR(z)}
\begin{split}
\partial_t \widetilde{R}_k&(z) =  \widetilde{R}_k(z) - 2(N-2) \widetilde{R}_k'(1) \widetilde{R}_k(z)- \dfrac{1}{2}(N-1)\\& \left[\widetilde{R}_k'(z)-2z\widetilde{R}_k'(1))\right]\widetilde{R}_k'(z) -  \dfrac{1}{2}(1-z^2)\big[-\widetilde{R}_k'(z)^2 \\&+ 2(\widetilde{R}_k'(1)-z\widetilde{R}_k'(z))\widetilde{R}_k''(z)+ (1-z^2)\widetilde{R}_k''(z)^2\big],
 \end{split}
\end{equation}
where $t=ln(k/\Lambda)$, and one also has 
\begin{equation}
  \label{eq_eta-etabar}
\eta_k = \epsilon \widetilde{R}_k'(1), \; \bar \eta_k = \epsilon [(N-1) \widetilde{R}_k'(1)-1].
\end{equation}

It is instructive to start by analyzing the flow equations for the first derivatives $\widetilde{R}_k'(z=1)$ and $\widetilde{R}_k''(z=1)$, assuming that $\widetilde{R}_k(z)$ is at least twice continuously differentiable
around $z=1$. these equations read: 
\begin{equation}
  \label{eq_beta_Rp1}
 -\partial_t \widetilde R_k'(1)=-\widetilde
  R_k'(1)+(N-2)\widetilde R_k'(1)^2 
\end{equation}
\begin{equation}
  \label{eq_beta_Rs1}
  \begin{split}
  -\partial_t \widetilde R_k''(1)=(-1+&6 \widetilde
  R_k'(1))\widetilde  R_k''(1)\\&+(N+7)\widetilde R_k''(1)^2+\widetilde  R_k'(1)^2.
      \end{split}
\end{equation}
If $\widetilde{R}_k(z)$ is analytic around $z=1$, the flow equations
for the higher derivatives evaluated in $z=1$ can be derived as
well. As noted by Fisher,\cite{fisher85} the expression for the $p$th
derivative only involves derivatives of lower or equal order. This
structure allows an iterative solution of the fixed-point equations
obtained by setting the left-hand sides to zero, provided of course
that $\widetilde{R}_*(z)$ has the required analytic property.

Beside the stable fixed point $\widetilde{R}_*'(1)=0$, there is one nontrivial fixed point associated with Eq.~(\ref{eq_beta_Rp1}): \begin{equation}
\label{eq_beta_Rp1FP}
\widetilde{R}_*'(1)=1/(N-2), 
\end{equation}
with a positive eigenvalue
$\Lambda_{1}=\epsilon$. This fixed point leads to the dimensional-reduction value of the critical exponents, \textit{i.e.},
$\eta=\bar{\eta}=\epsilon/(N-2)$, $\nu=1/\Lambda_{1}=1/\epsilon$. On the other hand, Eq.~(\ref{eq_beta_Rs1}) has nontrivial
fixed-point solutions only when $N\geq18$. These solutions are: 
\begin{equation}
\widetilde{R}_{\ast}''(1)=\frac{(N-8)+\sqrt{(N-2)(N-18)}}{2(N-2)(N+7)},
\end{equation}
which is unstable with an eigenvalue $\Lambda_{2}=\sqrt{\frac{(N-18)}{(N-2)}}\epsilon$, and
\begin{equation}
\label{eq_beta_Rs1stable}
\widetilde{R}_{\ast}''(1)=\frac{(N-8)-\sqrt{(N-2)(N-18)}}{2(N-2)(N+7)},
\end{equation}
which is stable with
$\Lambda_{2}=-\sqrt{\frac{(N-18)}{(N-2)}}\epsilon$. For $N<18$, no
fixed-point solutions exist for Eq.~(\ref{eq_beta_Rs1}). One instead
finds that there is a (finite) range of initial conditions
$\widetilde{R}_\Lambda'(1)$ for which the RG flow for
$\widetilde{R}_k''(1)$ leads to a divergence at a \textit{finite}
scale $k$, irrespective of the initial value
$\widetilde{R}_\Lambda''(1)$.

The solution to the absence of a nontrivial, twice differentiable
fixed-point function $\widetilde{R}_*(z)$ when $N<18$ is that the
proper fixed point controlling the critical behavior is nonanalytic
around $z=1$, with $\widetilde{R}_*'(z)$ having a cusp, \textit{i.e.},
a term proportional to $\sqrt{1-z}$ when $z\rightarrow1$. Numerical
solutions showing this cuspy behavior have been given by
Feldman\cite{feldman02} for $N=3, 4, 5$ and by us for general values
of $N<18$.\cite{tarjus04}

We have shown in detail in Ref.~[\onlinecite{tissier06b}] that the value
$N_{DR}=18$ separates a region in which $\widetilde{R}_*'(z)$ at the
critical, \ie, once unstable, fixed point has a cusp ($N<N_{DR}$) from
a region ($N>N_{DR}$) where $\widetilde{R}_{\ast}'(z)$ has only a
weaker nonanalyticity, a ``subcusp'' in $(1-z)^{\alpha(N)}$ with
$\alpha(N)$ a noninteger larger than $3/2$. The occurence of a cusp
changes the values of $\eta$ and $\bar{\eta}$ from the dimensional
reduction prediction,
$\eta_{DR}=\bar{\eta}_{DR}=\epsilon/(N-2)$. Indeed, the flow equation
for $\widetilde{R}'_k(1)$ is modified according to:
\begin{equation}
\begin{split}
  \label{eq_beta_Rp1cusp}
-\partial_t &\widetilde R_k'(1) =-\widetilde R_k'(1)+(N-2)\widetilde R_k'(1)^2 +\\& \lim_{z\rightarrow 1}\big\{ 2 (1-z)\widetilde R_k''(z)\left[ 2(1-z)\widetilde R_k'''(z)-3\widetilde R_k''(z)\right]+\\& \left[(N+1)\widetilde R_k''(z)-2(1-z)\widetilde R_k'''(z)\right](\widetilde R_k'(z)-\widetilde R_k'(1)) \big\} ,
\end{split}
\end{equation}
where the whole term $\lim_{z\rightarrow 1}(...)$ is nonzero when a
cusp is present in $\widetilde R_k'(z)$. As a result, the once
unstable fixed-point solution for $\widetilde{R}'_*(1)$ is no longer
equal to $1/(N-2)$ and it follows from Eqs. (\ref{eq_eta-etabar}) that
dimensional reduction is broken.\footnote{The correlation length
  exponent $\nu$ is still given, however, by its dimensional reduction
  value, $\nu=1/\epsilon$, but this is no longer true at the next
  ($2$-loop) order in $\epsilon$: see Ref. [\onlinecite{tissier06b}].}

\begin{figure}[htbp]
  \centering
  \includegraphics[width=.9 \linewidth]{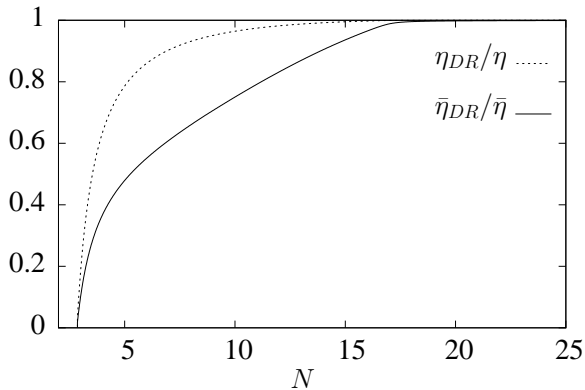}
  \caption{Breakdown of dimensional reduction near $d=4$: $\eta_{DR}/
    \eta$ (lower curve) and $\bar\eta_{DR}/ \bar\eta$ (upper
    curve) versus $N$ for the RF$O(N)$M with $N\geq 3$ at first order
    in $\epsilon=d-4$. The dimensional reduction value of the
    exponents is $\eta_{DR}=\bar\eta_{DR}=\epsilon/(N-2)$.}
  \label{fig_eta_etab}
\end{figure}

On the other hand, the weaker nonanalyticity occuring for $N>18$ does
not alter the flow equation for $\widetilde{R}_{k}'(1)$, which is
still given by Eq. (\ref{eq_beta_Rp1}), and dimensional reduction
still applies; in particular, $\eta=\bar{\eta}=\eta_{DR}$.  This
drastic change of behavior at $N=18$ is illustrated in
Figure~\ref{fig_eta_etab} where $\eta_{DR}/\eta$ and
$\bar{\eta}_{DR}/\bar{\eta}$ are plotted as a function of $N$.

To get some insight into the order of the nonanalyticity, one may
analyze the hierarchy of flow equations for the successive derivatives
of $\widetilde{R}_{k}(z)$ evaluated in $z=1$.\cite{fisher85,tissier06b} As
explained above, a fixed point with a well defined second derivative
and an associated negative eigenvalue ($\Lambda_{2}<0$) can be found
for $N>18$ (see Eq.~(\ref{eq_beta_Rs1stable})). \footnote{
  Cuspy fixed points are also present when $N>18$, but they are more
  than once unstable and correspond to possible multicritical
  behavior: see Refs.~[\onlinecite{tissier06b,sakamoto06}].}

\begin{figure}[htbp]
  \centering
  \includegraphics[width=\linewidth]{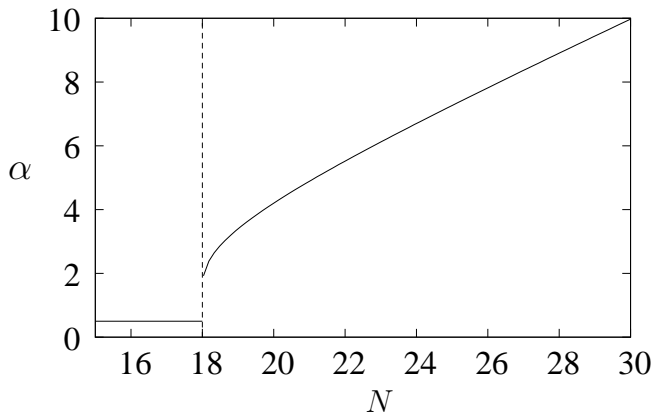}
  \caption{Exponent $\alpha(N)$ characterizing the order of the
    nonanalyticity in the cumulant of the renormalized random field,
    $R'(z)$, near $z=1$ for the RF$O(N)$M in $d=4+\epsilon$. A cusp
    leading to breakdown of dimensional reduction corresponds to
    $\alpha(N)=1/2$ and is obtained for $N<18$. There is a
    discontinuity at $N=18$ with $\alpha(N=18^+)=3/2$.}
  \label{fig_nonanalyticity}
\end{figure}

Let us assume that the first $p$ derivatives of $\widetilde{R}_{k}(z)$ are well defined in $z=1$. Contrary to the flow equations for $\widetilde{R}_{k}'(1)$ and $\widetilde{R}_{k}''(1)$ (see Eqs. (\ref{eq_beta_Rp1},\ref{eq_beta_Rs1})),  those for the
higher derivatives are linear, namely,
\begin{equation}
  \label{eq_flow_rp}
  \begin{split}
-\partial_t\widetilde R_k^{(p)}(1)=&\Lambda_p(\widetilde R_k'(1),
  \widetilde R_k''(1))\  \widetilde R_k^{(p)}(1)\\&+\mathcal F_p(\widetilde
  R_k'(1), \widetilde R_k''(1),\cdots, \widetilde R_k^{(p-1)}(1)),
    \end{split}
\end{equation}
where $\Lambda_p$ and $\mathcal F_p$ are known functions easily derived from Eq. (\ref{eq_tildeR(z)}). If $\widetilde{R}_k'(1)$ and $\widetilde{R}_k''(1)$ are chosen equal to
their fixed-point values given in Eqs. (\ref{eq_beta_Rp1FP},\ref{eq_beta_Rs1stable}),
one finds that
\begin{equation}
  \label{eq_eigenvalue_p}
  \begin{split}
\Lambda_{p*}&=\frac\epsilon{N-2}\Big[2p^2-(N-1)p+(N-2)+\\&\frac{p(N-5+6p)}
  {2(N+7)}(N-8-\sqrt{(N-2)(N-18)})\Big] .
   \end{split}
\end{equation}
For a given $N$, there exists an integer value $p_{\sharp}(N)$ such
that $\Lambda_{p*}<0$ for $p \leq p_{\sharp}(N)$ and $\Lambda_{p*}>0$
for $p\geq p_{\sharp}(N)+1$. The RG flow for the ($p_{\sharp}(N)+1$)th
derivative therefore diverges when $ t \rightarrow - \infty$ whereas
all lower-order derivatives reach finite fixed-point values.  As a
consequence, the fixed-point function $\widetilde{R}_{*}'(z)$ must
have a nonanalyticity of the form $(1-z)^{\alpha(N)}$ with
$p_{\sharp}(N)-1<\alpha(N)<p_{\sharp}(N)$. Refining the
reasoning,\cite{tissier06b} one finds that $\alpha(N)$ is given by the
solution of $\Lambda_{\alpha(N)+1*}=0$, where $\Lambda_{\alpha(N)+1*}$
is given by Eq. (\ref{eq_eigenvalue_p}) with $p$ replaced by the
noninteger $\alpha(N)+1$. The result is shown in Figure
\ref{fig_nonanalyticity}: the order of the non-analyticity increases
with $N$ when $N>18$, starting from $3/2$ when $N \rightarrow 18^+$,
and it goes as $N/2+O(1)$ at large $N$.

Finally, we stress the different ways in which ``cusp'' and
``subcusps`` appear along the RG flow. As seen above, subcusps occur
only at infinite RG time (\textit{i.e.}, at the fixed point). On the
contrary, due to the nonlinear nature of the beta function for
$\widetilde{R}_k''(1)$ (see Eq. (\ref{eq_beta_Rs1})), a cusp appears
at a finite time, which one may define as a ``Larkin scale'' by
analogy with the behavior of disordered elastic
systems.\cite{larkin70,larkin79,ledoussal02b,ledoussal04}

\subsection{Analytic versus nonanalytic behavior in the RFIM}

Consider now the Ising version of the random field model at zero
temperature. We have seen in section IV-B of paper I that
``anomalous'' contributions to the combination of running exponents
$2\eta_k-\bar \eta_k$ may appear if the dimensionless renormalized
cumulant $\delta_k(\varphi_1,\varphi_2)$ becomes nonanalytic as the
two field arguments approach each other, $\varphi_2 \rightarrow
\varphi_1$. This can be more conveniently studied by changing the
variables to $x=(\varphi_1+\varphi_2)/2$ and
$y=(\varphi_1-\varphi_2)/2$; the dependence of $\delta_k$ on $x$ is
anticipated as being completely regular and that on $y$ as potentially
anomalous. (Recall that due to the $Z_2$ and permutational symmetries,
$\delta_k$ is even in $x$ and $y$ separately.) The flow equation for
$\delta_k(x,y)$ is obtained from that for $v_k(\varphi_1,\varphi_2)$
given in paper I by deriving with respect to $\varphi_1$ and
$\varphi_2$ and switching to the new variables $x$ and $y$. One finds
that
\begin{widetext}
\begin{equation}
 \label{eq_delta_ising}
\begin{split}
  \partial_t \delta_k(x&,y)= (2\eta-\bar\eta) \delta(x,y) +
  \frac{1}{2}(d-4+\bar\eta) (x\partial_{x}+ y\partial_{y})
  \delta(x,y) - v_d \Big\{ \frac{1}{2} \Big[l_{2}^{(d)}(w_+)
  \delta_{0+}+ l_{2}^{(d)}(w_-)\delta_{0-}\\&- 2 
  l_{1,1}^{(d)}(w_+,w_-) \delta(x,y)\Big] \delta^{(02)}(x,y) -
  l_{1,1}^{(d)}(w_+,w_-) \delta^{(01)}(x,y)^2+ \Big[(
  l_{2}^{(d)}(w_+) \delta_{0+}' - l_{2}^{(d)}(w_-)\delta_{0-}' ) \\&+ 2(
  l_{2,1}^{(d)}(w_+,w_-)w_+' -  l_{1,2}^{(d)}(w_+,w_-)w_-')
  \delta(x,y)- 2( l_{3}^{(d)}(w_+)w_+' \delta_{0+}-
  l_{3}^{(d)}(w_-) w_-' \delta_{0-}) \Big]\delta^{(01)}(x,y)
  \\&+ \Big[ l_{2}^{(d)}(w_+) \delta_{0+}-
  l_{2}^{(d)}(w_-)\delta_{0-}\Big]\delta^{(11)}(x,y)+ 2
  l_{2,2}^{(d)}(w_+,w_-)w_+' w_-' \delta(x,y)^2 +
  l_{1,1}^{(d)}(w_+,w_-) \delta^{(10)}(x,y)^2  \\&+\Big[
  (l_{2}^{(d)}(w_+) \delta_{0+}'+ l_{2}^{(d)}(w_-)\delta_{0-}' ) - 2(
  l_{3}^{(d)}(w_+)w_+' \delta_{0+}+ l_{3}^{(d)}(w_-)w_-'
  \delta_{0-}) - 2 ( l_{2,1}^{(d)}(w_+,w_-)w_+' \\&+
  l_{1,2}^{(d)}(w_+,w_-) w_-' )  \delta(x,y) \Big]
  \delta^{(10)}(x,y) + \frac{1}{2} \Big[ l_{2}^{(d)}(w_+) \delta_{0+}+
  l_{2}^{(d)}(w_-)\delta_{0-}+  2
  l_{1,1}^{(d)}(w_+,w_-)\delta(x,y)\Big] \delta^{(20)}(x,y) \Big\},
\end{split}
\end{equation}
  
\end{widetext}
where we have dropped the subscript $k$ in the right-hand side and
introduced the short-hand notation $w_\pm=u_k''(x\pm y)$,
$\delta_{0\pm}=\delta_{k,0}(x\pm y)$ with $\delta_{k,0}(x)=\delta_k(x,
y=0)$; for a function of a single argument a prime denotes a
derivative whereas for functions of two arguments partial derivatives
are indicated as superscripts, \textit{e.g.},
$\delta^{(10)}(x,y)=\partial_x \delta(x,y)$,
$\delta^{(01)}(x,y)=\partial_y \delta(x,y)$, etc. Finally,
$l_{n}^{(d)}(w)$ and $l_{n_1,n_2}^{(d)}(w_1,w_2)$ are the
``dimensionless threshold functions'' defined from the infrared cutoff
function $\widehat{R}_k(q^2)=Z_k q^2 r(q^2/k^2)$ (see paper I and
Ref. [\onlinecite{berges02}]).

We now follow a reasoning similar to that developed for the RF$O(N)$M
near $d=4$. Assume that $\delta_k(x,y)$ is continuously differentiable
with respect to $y$ around $y=0$ up to some order $2p$ with $p\geq
1$. Then, introducing the notation $\delta_{k,q}(x)=\partial_y^q
\delta_k(x,y)\vert_{y=0}$ and using the property that all derivatives
of odd order vanish in $y=0$ due to the inversion symmetry, one may
express $\delta_k(x,y)$ in the vicinity of $y=0$ as
\begin{equation}
 \delta_k(x,y)=\sum_{q=0}^{p}\dfrac{y^{2q}}{(2q)!}\delta_{k,2q}(x) + o(y^{2p}).
\end{equation}
By inserting this expression in the RG flow equation for $\delta_k(x,y)$, Eq. (\ref{eq_delta_ising}), one derives the following flow equations for the function evaluated in $y=0$,
\begin{equation}
 \label{eq_delta0_ising}
\begin{split}
&\partial_t \delta_{k,0}(x)=(2\eta_k-\bar\eta_k) \delta_{k,0}(x) +  \frac{1}{2}(d-4+\bar\eta_k) x\delta_{k,0}'(x) - \\& 2v_d \Big\{ l_{4}^{(d)}(u''_k(x))  \delta_{k,0}(x)^2 u'''_k(x)^3- 4 l_{3}^{(d)}(u''_k(x))\; u'''_k(x) \times \\& \delta_{k,0}(x) \delta_{k,0}'(x)+ l_{2}^{(d)}(u''_k(x)) [\frac{3}{2}\delta_{k,0}'(x)^2 + \delta_{k,0}(x) \delta_{k,0}''(x)]\Big\},
\end{split}
\end{equation}
and for the derivatives,
\begin{equation}
\begin{split} 
\label{eq_delta2_ising}
\partial_t \delta_{k,2}(x) = -L_{2}[u_k'',\delta_{k,0}] \delta_{k,2}(x)& + 3v_d l_2^{(d)}(u_k''(x)) \delta_{k,2}(x)^2\\& -2v_d \mathcal G_{2}[u_k'',\delta_{k,0}],
\end{split}
\end{equation}
and for $p\geq2$,
\begin{equation}
\begin{split}
\label{eq_delta2p_ising} 
\partial_t \delta_{k,2p}(x) = - L_{2p}[u_k'',&\delta_{k,0},\delta_{k,2}] \delta_{k,2p}(x) \\&-2v_d \mathcal G_{2p}[u_k'',\left\lbrace \delta_{k,2q}\right\rbrace _{q \leq p-1}],
\end{split}
\end{equation}
where $L_{2}[u_k'',\delta_{k,0}]$ and $L_{2p}[u_k'',\delta_{k,0},\delta_{k,2}]$ are linear operators whose expressions are given in Appendix A. The $\mathcal G_{2p}$'s are functionals of $u_k''(x)$,  $\delta_{k,0}(x)$, and of the derivatives $\delta_{k,2q}(x)$ with $q \leq p-1$. Their expressions are not worth displaying here.

The above equations are complemented by the flow equation for $u_k(x)$, or its derivative $u'_k(x)$, obtained from the results of section IV-B in paper I,
\begin{equation}
\begin{split}
  \label{eq_u'_ising}
\partial_t u_k'& (x)=-(2-\eta_k)u_k'(x) + \frac{1}{2}(d-4+\bar\eta_k) x u_k''(x) + \\&2v_d\Big\{l_1^{(d)}(u''_k(x))\delta_{k,0}'(x)- l_2^{(d)}( u''_k(x))u'''_k(x)\delta_{k,0}(x)\Big\},
\end{split}
\end{equation}
and the expression for the running anomalous dimension,\cite{tarjus07_1}
\begin{equation}
\begin{split}
  \label{eq_eta_ising}
\eta_k= \frac{8v_d}{d} & \Big\{2 m_{3,2}^{(d)} (u_{k}''(x_{m,k}),u_{k}''(x_{m,k}))u_{k}'''(x_{m,k})^2- \\&
m_{2,2}^{(d)} (u_{k}''(x_{m,k}),u_{k}''(x_{m,k}))u_{k}'''(x_{m,k})\Big\},
\end{split}
\end{equation}
where $x_{m,k}$ denotes the nontrivial configuration that minimizes
the $1$-replica potential, and which therefore satisfies
$u'_k(x_{m,k})=0$; the $m_{n_1,n_2}^{(d)}(w_1,w_2)$'s are additional
``dimensionless threshold functions'' (see paper I and
Ref.~[\onlinecite{berges02}]). Similarly, one also has an expression
for $2\eta_k- \bar \eta_k$ which is derived from
Eqs. (\ref{eq_delta0_ising}) and (\ref{eq_u'_ising}) and the
constraint $\delta_{k,0}(x_{m,k})=1$ (one recovers the equation of
paper I, but now without the anomalous terms).

One first notices that the RG equations for $u'_k(x)$, $\delta_{k,0}(x)$, and $\eta_k$ form a closed set. No aditional input is required from the derivatives $\delta_{k,2p}(x)$ with $p\geq  1$, which means that the RG flow for the $1$-replica potential and for the $2$-replica potential (or the second cumulant) evaluated for equal field arguments is closed without further knowledge of the full field dependence of the $2$-replica potential for distinct replicas. This property is a direct consequence of the assumptation that the behavior of the second cumulant is sufficiently regular when the two field arguments become equal, more precisely, that $\partial_y^2 \delta_k(x,y)$ is finite when $y\rightarrow 0$.

Before discussing the consequences of this property, it is worth
mentioning that it results from the structure of the exact RG
equations and not from the specific approximation chosen here. More
generally indeed, the exact RG flows for the $1$-replica component of
the effective average action and for the cumulants of the renormalized
random field (see sections II-C,D of paper I) evaluated for equal
field arguments decouple from the full functional dependence of the
cumulants when the latter is regular enough in the limit of equal
arguments. This point will be further developed and clarified in a
forthcoming publication centered on the superfield
formalism.\cite{tarjus07_susy}

As in the previously discussed case of the RF$O(N)$M near $d=4$, one
expects that the fixed point obtained without reference to distinct
replicas, and associated with a regular enough behavior of the
cumulants in the limit of equal arguments, corresponds to dimensional
reduction. To prove this, one needs to show that it is equivalent, in
the $1$-replica sector at least, to the corresponding fixed point of
the pure system in two dimensions less. This is indeed illustrated
near the upper critical dimension $d=6$: it is easy to show (see also
section IV-D of paper I) that, at first order in $\epsilon=6-d$,
Eqs. (\ref{eq_delta0_ising}, \ref{eq_u'_ising}, \ref{eq_eta_ising})
give back the result of the pure Ising model at first order in
$\epsilon=4-d$. The difficulty in going beyond this step is that the
present truncation does not necessarily preserve the underlying
supersymmetry of the model. For instance, the second order in
$\epsilon=6-d$ of Eqs. (\ref{eq_delta0_ising}, \ref{eq_u'_ising},
\ref{eq_eta_ising}) breaks the dimensional reduction property;
however, this is clearly an artefact of the truncation and of the
  choice of regulator. One can check this by improving the treatment
so that the exact $2$-loop results are recovered near $d=6$: the
calculation becomes extremely tedious in the present formalism and a
nonperturbative closure becomes hardly tractable numerically; but it
is nonetheless found by a direct analysis near $d=6$ that if the $2$-
and $3$-replica cumulants are regular enough in their field dependence
so that the flow equations evaluated for equal replica fields
decouples as in Eqs. (\ref{eq_delta0_ising}, \ref{eq_u'_ising},
\ref{eq_eta_ising}), the corresponding fixed point at second order in
$\epsilon$ leads to dimensional reduction.

Awaiting for a proper resolution of the problem via the superfield
formalism,\cite{tarjus07_susy} we will associate with dimensional
reduction the fixed point corresponding to
Eqs. (\ref{eq_delta0_ising}, \ref{eq_u'_ising}, \ref{eq_eta_ising}),
fixed point that can be continuously followed as a function of
dimension $d$ and, as will be discussed below, as a function of the
number of components $N$. Breaking of dimensional reduction therefore
implies the occurence of a strong enough nonanalyticity in the field
dependence of the renormalized cumulants of the disorder. ``Strong
enough'' here means that it is sufficient to couple the flow of the
components of the effective average action evaluated for equal fields
to the full functional dependence involving distinct replica fields.

From the flow equation for $\delta_k(\varphi_1,\varphi_2)\equiv
\delta_k(x,y)$, Eq. (\ref{eq_delta_ising}), it is clear that the only
way to avoid dimensional reduction is therefore the existence of a
linear cusp in the fixed-point function $\delta_*(x,y)$,
\textit{i.e.}, with
\begin{equation}
 \label{eq_delta_cusp}
\delta_*(x,y)= \delta_{*,0}(x)  +  \vert y \vert \delta_{*,a}(x) + O(y^2),
\end{equation}
in the vicinity of $y=0$. This leads to the appearance of an
``anomalous'' contribution to the expression of the beta function for
$\delta_{k,0}(x)$:
\begin{equation}
 \label{eq_beta_delta0_cusp}
\beta_{\delta_0}(x)= \beta_{\delta_0}\vert_{reg}(x)- v_d l_2^{(d)}(u''_*(x))\delta_{*,a}(x)^2,
\end{equation}
where $-\beta_{\delta_0}\vert_{reg}(x)$ is given by the right-hand
side of Eq. (\ref{eq_delta0_ising}). The above equation gives back the
expression of $2\eta_k- \bar \eta_k$ derived in section IV-B of paper
I with the temperature set to zero.

Before closing this discussion of the RFIM, we would like to emphasize
a few additional points which parallel the comments made in the
previous subsection. First, the appearance of a cusp in
$\delta_k(x,y)$ is associated with the divergence of the second
derivative $\delta_{k,2}(x)$. Due to the nonlinear character of the
flow equation for $\delta_{k,2}(x)$, Eq. (\ref{eq_delta2_ising}), this
divergence, if present, is expected to first occur at a finite scale
which, as before, we generically call the ``Larkin scale''. For a
running scale $k$ larger than the Larkin scale $k_L$, the effective
average action is analytic and it develops a cusp (in $\delta_k(x,y)$)
for $k$ less than $k_L$. From the flow equation for $\delta_k(x,y)$,
one can see that a linear cusp is stable under RG flow, in that it
does not lead to stronger ``supercusps''.

Secondly, the appearance of a ``subcusp'' in $\delta_k(x,y)$ is
signaled by the divergence of a higher-order derivative
$\delta_{k,2p}(x)$, with $p \geq 2$ being related to the order of the
nonanalyticity (which is strictly less than $2p$ and strictly more
than $2(p-1)$). The flow equation for $\delta_{k,2p}(x)$ being linear,
a subcusp can only appear at infinite RG time, \textit{i.e.}, at the
fixed point. Following the reasoning developed in Appendix A, we
conclude that the order of the nonanalyticity characterizing the fixed
point increases as $1/(6-d)^2$ when $d$ approaches $6$ from
below. This is the counterpart of the situation found near $d=4$ where
the order of the nonanalyticity increases as $N/2$ when the number of
components $N$ gets large. In both cases the critical fixed points
with fully analytic (dimensionless) effective action found above $d=6$
and when $N\rightarrow \infty$ are approached in $d$ or $N$ by fixed
points with diverging orders of the nonanalyticity, \textit{i.e.},
with weaker and weaker subcusps. We recall however that such subcusps
are not sufficient to break dimensional reduction.

\subsection{Extension to the whole $(N,d)$ plane}

The preceding developments on the connection between breakdown of
dimensional reduction and nonanalyticity of the effective average
action can be extended to the whole $(N,d)$ plane. The nonanalyticity
now occurs in the renormalized cumulants of the disorder as two
arguments, \textit{i.e.}, two replica fields, approach each other,
$\vect \varphi_2 \rightarrow \vect \varphi_1$. If the nonanalyticity
is weak enough, namely, if it is weaker than a linear cusp in the
second cumulant of the renormalized random field, the RG flows for
$u_k(\rho), \delta_{k,T}(\rho),\delta _{k,L}(\rho), \eta_k$ (see
section IV-C of paper I) decouple from those involving distinct
replica fields. The associated fixed point can be continuously
followed in the $(N,d)$ plane and, near $d=6$, near $d=4$ for $N>18$
and when $N \rightarrow \infty$, it leads to dimensional
reduction. For reasons explained above, a direct proof that it
corresponds to dimensional reduction away from the perturbative regime
is hampered by the truncation used here, but we rely on the continuity
argument within the $(N,d)$ plane to nonetheless identify it with an
approximation of the dimensional reduction fixed point.

Breaking of dimensional reduction is thus associated with the
appearance of a sufficiently strong nonanalyticity in the $2$-replica
potential $v_k(\rho_1,\rho_2,z)$. Analysis of the flow equation for
$v_k(\rho_1,\rho_2,z)$ (see paper I) shows that this corresponds to
the following behavior as $\vect \varphi_2 \rightarrow \vect
\varphi_1$:
\begin{equation}
 \label{eq_cusp_v_O(N)}
 v_k(\rho_1,\rho_2,z)\simeq v_{k,reg}(\rho_1,\rho_2,s^2) + 
|s|^3 v_{k,a}(\rho,r^2,|s|),
\end{equation}
with $s^2=\vert \vect \varphi_1 - \vect
\varphi_2\vert^2/(8\sqrt{\rho_1\rho_2})$, $\rho=(\rho_1+\rho_2)/2$,
$r=(\rho_1-\rho_2)/(4|s|)$, and $\vert \vect \varphi_1 - \vect
\varphi_2\vert^2=2(\rho_1+\rho_2-2\sqrt{\rho_1\rho_2}z)$; $v_{k,reg}$
and $v_{k,a}$ are analytic functions of their arguments in the
vicinity of $\rho_2=\rho_1=\rho$, $z=1$ (\textit{i.e.}, $s=0$), and
$r^2\lesssim \rho^2$. The cusp in the second cumulant of the
renormalized random field,
$\delta_k^{\mu\nu}(\rho_1,\rho_2,z)= \partial
_{\varphi_1^\mu} \partial_{ \varphi_2^\nu} v_k(\rho_1,\rho_2,z)$,
occurs in $|s|$ or, equivalently, in $\vert \vect \varphi_1 - \vect
\varphi_2\vert$; it is marked by the divergence of the second
derivative of $v_k$ with respect to $s^2$ when $s=0$ (which also
implies $\rho_2=\rho_1$). Note however that on top of $\vert \vect
\varphi_1 - \vect \varphi_2\vert$, there is now an additional
variable, denoted $r$ above, that characterizes the way $\vect
\varphi_2$ approaches $\vect \varphi_1$.

To conclude this section, it should be stressed that the consistency of the present scenario and the actual occurence of a cusp in a region of the $(N,d)$ diagram must be verified by a numerical resolution of the NP-FRG equations. This is what we address now.

\section{A unified description of criticality, ferromagnetism and QLRO}
\label{sec_unified}

\subsection{RG flow equations and their numerical resolution}

The RG flow equations for the RF$O(N)$M in the minimal truncation of the NP-FRG discussed above are given in paper I. Focusing on the fixed points and their vicinity, we drop the subdominant terms involving the temperature (it will be checked that the temperature exponent $\theta$ is indeed strictly positive). The structure of the resulting equations can be summarized as follows:
\begin{equation}
\label{eq_beta_u'_O(N)}
\partial_t u'_k(\rho) = -\beta_{u'}[u_k',\delta_{k,T},\delta_{k,L};\eta_k,\bar \eta_k] (\rho),
\end{equation}
\begin{equation}
\label{eq_beta_v_O(N)}
\partial_t  v_k(\rho_1,\rho_2,z) = -\beta_{v}[u_k',v_k;\eta_k,\bar \eta_k] (\rho_1,\rho_2,z),
\end{equation}
\begin{equation}
\label{eq_beta_eta_O(N)}
\eta_k= \gamma_{\eta}(\rho_{m,k},u''_k(\rho_{m,k}),\delta_{k,T}(\rho_{m,k}),\delta_{k,L}(\rho_{m,k})),
\end{equation}
where $\beta_{u'}$ and $\beta_v$ are functionals and $\gamma_{\eta}$
is a function; $\delta_{k,T}(\rho)=(2\rho)^{-1}\partial_z
v_k(\rho,\rho,z)\vert_{z=1}$ and
$\delta_{k,L}(\rho)=2\rho \partial_{\rho_1} \partial_{\rho_2}
v_k(\rho_1,\rho_2,z=1)\vert_{\rho_1=\rho_2=\rho}$ are the transverse
and longitudinal components of the second cumulant of the renormalized
random field evaluated for equal field arguments, and $\rho_{m,k}$ is
the configuration that minimizes the $1$-replica potential
($u_k'(\rho_{m,k})=0$). The running exponent $\bar \eta_k$ is derived
from the flow of the constraint $\delta_{k,T}(\rho_{m,k})=1$ that
follows from the definition of the renormalized
temperature.\cite{tarjus07_1}

Eqs. (\ref{eq_beta_u'_O(N)}-\ref{eq_beta_eta_O(N)}) form a set of
coupled partial differential equations involving in particular a
function of three variables. Studying the whole $(N,d)$ plane by
numerically solving these equations remains a very difficult and
computationally intensive task. To facilitate the study, we have used
in addition an expansion in powers of the fields. However, some
caution must be exerted in order to retain enough of the functional
character for allowing a description of possible cusp or nonanalytic
dependence. We have therefore considered an expansion of
$\rho_1,\rho_2$ around the configuration $\rho_{m,k}$ while keeping
the complete dependence on the variable $\vert \vect \varphi_1 - \vect
\varphi_2\vert^2$ which we anticipate to be the key variable for
describing the cusp (see above). More specifically, we have chosen the
following approximation:
\begin{equation}
u_k(\rho)=\left(\frac{\lambda_k}{8} \right) (\rho-\rho_{m,k})^2,
\end{equation}
\begin{equation}
\begin{split}
v_k(&\rho_1,\rho_2,z)=v_{k,00}(s^2)+v_{k,10}(s^2)(\rho_1+\rho_2-2 \rho_{m,k})+
\\& \frac{1}{2}v_{k,20}(s^2)(\rho_1+\rho_2-2 \rho_{m,k})^2+\frac{1}{2} v_{k,02}(s^2)(\rho_1-\rho_2)^2,
\end{split}
\end{equation}
where $s^2=\vert \vect \varphi_1 - \vect
\varphi_2\vert^2/(8\rho_{m,k})=(\rho_1+\rho_2-2\sqrt{\rho_1\rho_2}z)/(4\rho_{m,k})$
(when $\rho_1=\rho_2=\rho_{m,k}$, $s^2$ varies then between $0$ and
$1$); by construction, $v_{k,00}'(0)=\delta_{k,T}(\rho_{m,k})=1$. It
is easily checked that this truncation still reproduces the
perturbative results near $d=6$, $d=4$, and when $N\rightarrow \infty$
(compare with section V of paper I).

Inserting the above expressions into the flow equations,
Eqs. (\ref{eq_beta_u'_O(N)}-\ref{eq_beta_eta_O(N)}), provides a set of
coupled partial differential equations for $4$ functions,
$v_{k,00}(s^2), v_{k,01}(s^2), v_{k,20}(s^2), v_{k,02}(s^2)$, and $3$ running
parameters, $\rho_{m,k}, \lambda_k, \eta_k$ (plus, when convenient,
$\bar \eta_k$), whose resolution now represents a more tractable
numerical problem.

We close this subsection by outlining the numerical methods used for
solving the partial differential equations. For each couple $(N,d)$
and for a choice of the infrared cutoff function (in most
calculations, we have taken the ``optimized'' regulator\cite{litim00}
described in paper I that leads to explicit analytic expressions for
all the dimensionless threshold functions appearing in the beta
functions), we follow the evolution under RG flow of the various
functions and parameters for given initial conditions. For the
functions, a finite difference mesh is used for the variable $s$, so
that standard algorithms are sufficient to solve the evolution with
the RG ``time'' $t$ (or equivalently, the scale $k$). When necessary,
most notably for checking the robustness of a nonanalytic cusp-like
behavior around $s=0$, we vary the mesh spacing. To reach the
critical, once unstable, fixed point, we fine tune the initial
condition for $\rho_{m,k}$ which represents the unstable direction. We
consider that a fixed point is attained when the sum of the absolute
values of all beta functions is less than $10^{-6}$. The whole
procedure can be accelerated by following the fixed points by
continuity (when possible) in the $(N,d)$ plane through small finite
changes of $N$ and/or $d$.

We now move on to the presentation of the main results.

\subsection{Dimensional reduction and its breaking}

To study the ``weakly nonanalytic'' critical fixed point associated
with dimensional reduction (see section III), one does not need the
full dependence on $s$ of the functions $v_{k,00}, v_{k,01}, v_{k,20},
v_{k,02}$, but only their value and that of their first derivative
evaluated in $s=0$. In addition, to check the stability of this fixed
point to the appearance of a cusp in $|s|$ in the dimensionless
cumulant of the renormalized random field $\delta_k$ (or,
equivalently, a term in $|s|^3$ in $v_k$), we also follow the second
derivative of the functions, evaluated in $s=0$.

\begin{figure}[htbp]
  \centering
  \includegraphics[width=\linewidth]{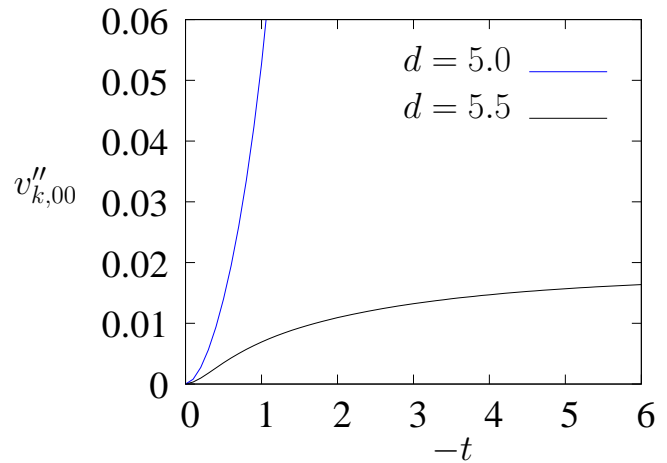}
  \caption{ Difference in behavior above and below $d_{DR}(N)$ for
    $N=3$: evolution with RG ``time'' $ | t |$ of $v_{k,00}''(0)$
    which, up to a constant prefactor, is the second derivative of the
    dimensionless $2$-replica potential $v_k$ with respect to
    $ s^2 \propto |\mathbf{\varphi}_1-\mathbf{\varphi}_2|^2$ evaluated in
    $\rho_1=\rho_2=\rho_{m,k}$ and $s=0$. The initial condition on
    $\rho_{m,k}$ has been fine tuned so that the other running
    quantities reach the dimensional reduction fixed point. For
    $d=5.5$, $v_{k,00}''(0)$ reaches a finite fixed-point value;
    conversely, for $d=5.0$, it diverges at a finite ``Larkin'' time.
  }
\label{fig_divergence}
\end{figure}

We find that the dimensional reduction fixed point is stable versus
cusp-like behavior in a whole region of the $(N,d)$ plane. However,
there is a critical dimension $d_{DR}(N)$ depending on $N$ (which one
may as well describe for fixed $d$, as a critical number of components
$N_{DR}(d)$) at which the second derivative of $v_k$ with respect to
$s^2$ in $s=0$ first diverges along the RG flow at a finite ``Larkin''
scale. The difference in behavior above and below $d_{DR}(N)$ is
illustrated in Figure \ref{fig_divergence}. For $N=3$, we display the
evolution with $ t$ of the second derivative of $v_k$ with respect to
$s^2$ evaluated in $\rho_1=\rho_2=\rho_{m,k}$ and $s=0$, \textit{i.e.},
up to a constant prefactor, $v_{k,00}''(0)$. The initial condition on
$\rho_{m,k}$ has been fine tuned so that the other running quantities
(see above) reach the dimensional reduction fixed point. For $d=5.5$,
$v_{k,00}''(0)$ reaches a finite fixed-point value; conversely, for
$d=5.0$, it diverges at a finite Larkin time. (Note that the other
second derivatives $v_{k,10}''(0), v_{k,20}''(0), v_{k,02}''(0)$ all
diverge at the same Larkin scale.) The value of $d_{DR}(N)$ in this
case in about $5.1$.

\begin{figure}[htbp]
  \centering
  \includegraphics[width=\linewidth]{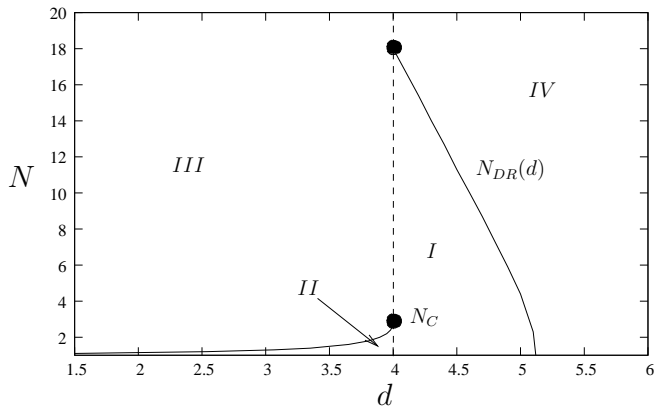}
  \caption{Predicted phase diagram of the $d$-dimensional
    RF$O(N)$M. In region $III$, there are no phase transitions and the
    system is always disordered (paramagnetic). In regions $I$ and
    $IV$, there is a second-order paramagnetic to ferromagnetic
    transition and in region $II$, a second-order transition between
    paramagnetic and QLRO phases.  In region $IV$ the nonanalyticity
    of the dimensionless effective action at the zero-temperature
    fixed point is weak enough to let the critical exponents take
    their dimensional reduction value, whereas a complete breakdown of
    dimensional reduction occurs in regions $I$ and $II$.  }
  \label{fig_diag_n_d}
\end{figure}

When $d\rightarrow 4^+$, we numerically recover the value $N_{DR}=18$,
thereby confirming that the nonperturbative truncation actually leads
back to the perturbative FRG result near $d=4$. The curve $N_{DR}(d)$
extends continuously down to $N=1$, where we obtain $d_{DR}(N)\simeq
5$. It separates the two regions denoted $I$ and $IV$ in
Figure~\ref{fig_diag_n_d}.

\begin{figure}[htbp]
  \centering
  \includegraphics[width=\linewidth]{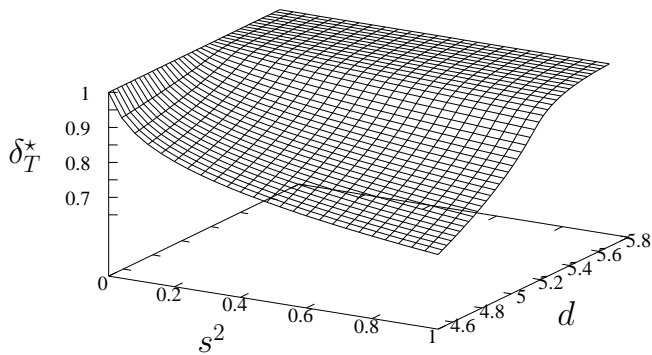}
  \caption{Presence or absence of a cusp in the fixed-point
    function $\delta_{*T}(\rho_{m,k},s^2)$ for $N=3$: $2$-dimensional
    plot of $\delta_{*T}(\rho_{m,k},s^2)$ versus $s^2$ and $d$. Below some
    dimension around $5$, a cusp is clearly visible near $s=0$.  }
  \label{fig_cusp_n_y}
\end{figure}

For $d<d_{DR}(N)$, the fixed point controlling the critical behavior
of the RF$O(N)$M must now be studied by keeping the full dependence on
$s$ of the renormalized disorder cumulant. We do find in this region
that a cusp occurs at a finite time that corresponds to the Larkin
scale discussed above. In Figure~\ref{fig_cusp_n_y} we illustrate the change of
behavior related to the presence or absence of cusp in the fixed-point
function $\delta_{*T}(\rho_{m,k},s^2)$ for $N=3$. We plot
$\delta_{*T}(\rho_{m,k},s^2)$ as a function of both $s^2$ and $d$. 
Below some dimension $5.1$, a cusp is clearly visible near $s=0$. We
have checked its robustness and that it is indeed a behavior in
$|s|$ by changing the spacing of the discretization of the $s$
variable. The value of $d$ for which the cusp first appears coincide
with $d_{DR}$ as determined previously from the divergence of
$\partial_{s^2}^2 v_k\vert_{\rho=\rho_{m,k},s=0}$; here,
$d_{DR}(N=3)\simeq 5.1$.

Finally, we draw attention to the fact that the critical exponents
evolve continuously upon crossing the critical line $d_{DR}(N)$. This
has been clearly illustrated in Figure~\ref{fig_nonanalyticity} for
the exponents $\eta$ and $\bar\eta$ near $d=4$: one can see that
$\bar\eta$ gradually separates from $\eta$ as $N$ moves down from $18$
and does \textit{not} settle in general to the value $\bar\eta=2\eta$.
This provides strong evidence in favor of a characterization of the
critical scaling behavior by three independent exponents and not two
as suggested in Refs. [\onlinecite{schwartz85b}].

\subsection{Criticality, ferromagnetism, and QLRO}

In the present NP-FRG formalism, the equilibrium phases of the system
are characterized by the flow evolution of the dimensionless order
parameter at scale $k$, $\rho_{m,k}$. Depending on the initial
conditions, $\rho_{m,k}$ is found to: 
\begin{enumerate}
\item diverge when time goes to $-\infty$ ($k\rightarrow 0$) in such a
  way that the dimensionfull order parameter (the magnetization) tends
  to a finite value; this behavior can be associated with a fully
  stable (attractive) fixed point describing long-range ferromagnetic
  order,
\item go to zero at a finite scale, which corresponds to the
  disordered paramagnetic phase, \footnote{In the present truncation
    that relies on an expansion around a nontrivial minimum,
    \textit{i.e.}, different from zero, we do not pursue the study of
    the flow once $\rho_{m,k}$ reaches zero; if one is interested in
    investigating the disordered phase, it is not difficult to change
    the truncation: see Ref. [\onlinecite{tetradis94}]. Note also that
    generically the disordered fixed points are not at zero
    temperature.}
\item reach a nontrivial fixed point value; the dimensionful order
  parameter which goes as $k^{d-4+\bar\eta}\rho_{m*}$ is then zero
  when $k=0$.
\end{enumerate}
 In the latter case, if only one
parameter (say, the initial value of $\rho_{m,k}$) need be adjusted,
the fixed point is once unstable and describes the critical point of
the model and the associated scaling behavior. This is the situation
found in regions $I$ and $IV$ of the $(N,d)$ diagram shown in
Figure~\ref{fig_diag_n_d} with, as discussed above, a qualitative
difference with respect to dimensional reduction between the two
regions.

A different pattern is found below dimension $4$ for models with
continuous symmetry ($N>1$). Whatever the initial conditions, the RG
flow no longer leads to a divergence of $\rho_{m,k}$, which is in line
with the predicted absence of long-range order in the RF$O(N)$M for
$d<4$. One obtains instead a nontrivial attractive fixed point
characterized by a cusp in the second cumulant of the renormalized
random field. This fixed point comes on top of the once unstable
(critical) fixed point also characterized by a cusp. It describes a
whole low-disorder phase that is associated with a nontrivial scaling
behavior. This phase has a vanishing order parameter but algebraically
decaying correlation functions characterized by the two anomalous
dimensions $\eta$ and $\bar\eta$ (see section IV-A of paper I). It
therefore corresponds to a QLRO phase and it transforms into the
disordered paramagnetic phase at a critical point, itself controlled
by a ``cuspy'' fixed point.

The QLRO phase only occurs below a critical value of the number of
components, $N_c=2.83...$, which can also be directly computed from an
analysis of the perturabative FRG equations at and near
$d=4$.\cite{ledoussal06,tissier06,tissier06b} It is very similar to
the QLRO phase found in elastic systems pinned by
disorder,\cite{giamarchi94,giamarchi95,giamarchi98} and for the
RF$XY$M ($N=2$) it actually identifies to the latter phase when
$d\rightarrow 4^-$. (The numerical solution of the truncated NP-FRG
equations then reduces, as expected, to the $1$-loop perturbative FRG
result for which the equivalence with a random periodic elastic model
is easily shown: see \textit{e.g.} paper I.)  However, contrary to the
situation in disordered elastic systems, the QLRO phase only exists
below a (bare) critical disorder and the existence of two ``cuspy''
fixed points provides a mechanism for destroying QLRO below some lower
critical dimension.\cite{ledoussal06,tissier06,tissier06b}

The variation with $N$ and $d$ of the characteristics of the two
``cuspy'' fixed points is illustrated in Figure 6 where we plot the
value of the anomalous dimension $\eta$ versus $d$ for a series of
values of $N$. For $N> N_c\simeq 2.83$, only one fixed-point value
emerges from the point $(\eta=0, d=4)$; but for $N<N_c$, one finds two
values of $\eta$ for each dimension, the upper one being associated
with the critical fixed point and the lower one with the QLRO fixed
point. One can see that the two fixed-point branches coalesce for a
value $d_{lc}(N)$ which consequently determines the lower critical
dimension below which no phase transition is observed.
\begin{figure}[htbp]
  \centering
  \includegraphics[width=\linewidth]{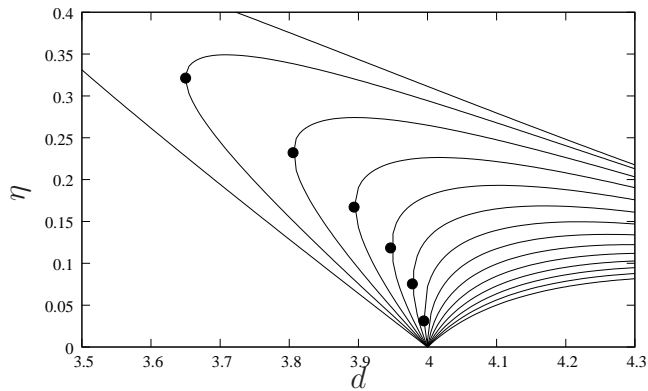}
  \caption{Anomalous dimension $\eta$ associated with the single or
    the two zero-temperature ``cuspy'' fixed points plotted versus $d$
    for values of $N$ ranging from 1.4 to 4 by steps of 0.2. For $N>
    N_c\simeq 2.83$, only one fixed-point value emerges from the point
    $(\eta=0, d=4)$; but for $N<N_c$, one finds two values of
    $\eta$ for each dimension, the upper one being associated with the
    critical fixed point and the lower one with the QLRO fixed
    point. The two branches of fixed points coalesce for a value
    $d_{lc}(N)$ shown by filled circles.  }
  \label{fig_eta_qlro}
\end{figure}

The lower critical dimension is shown in Figure~\ref{fig_diag_n_d},
where it separates region $II$ with QLRO and region $III$ with no
phase transition. Note that the result is compatible with Feldman's
prediction\cite{feldman00} that there is no QLRO above $N=3$. The
lower critical curve, seen as $N_{lc}(d)$ in place of $d_{lc}(N)$,
decreases as $d$ decreases and is expected to reach $N_{lc}=1$ for
$d=2$, which corresponds to the lower critical dimension (for
long-range ferromagnetism) in the Ising version. (In our present
approximate description, we find $N_{lc}(d=2)\simeq 1.15$ instead of
the exact result $N_{lc}=1$: see the discussion below.)  This behavior
is in fact reminiscent of what occurs in the pure $O(N)$
model.\cite{tissier06} Although never acknowledged, the $(N,d)$ phase
diagram of the latter, derived from the solution of phenomenological
RG equations\cite{cardy80} and from known exact results, is indeed
very similar to that of the RF$O(N)$M displayed in
Figure~\ref{fig_diag_n_d}: the critical value $N_c$ is now equal to
$2$ and occurs in $d=2$, which is the lower critical dimension for
ferromagnetism; a QLRO phase exists in a region between $(N=1,d=1)$
and $(N=2,d=2)$, which is the counterpart of region $II$ described
above, but is of course not physically relevant for real systems. This
QLRO is different from that associated with the low-temperature phase
of the $XY$ model in $d=2$, below the Kosterlitz-Thouless
transition,\cite{kosterlitz73,berezinskii70} as the latter corresponds
to a line of fixed points while the former is controlled by a single
fixed point.

The situation found in the RF$O(N>1)$M below $d=4$ raises two
questions. First, is a QLRO phase observable in an experimentally
realizable random field system ? The question is especially acute
because it has been suggested that the $3-d$ RF$XY$M displays such a
phase,\cite{giamarchi94,giamarchi95,gingras96} called ``Bragg glass''
in relation with pinned vortex lattices in disordered type-$II$
superconductors. the second question is more academic and concerns the
status of the singular point $(N_c=2.83...,d=4)$. Following the
analogy with its counterpart, $(N_c=2,d=2)$, in the pure $O(N)$ model
(see above), one may wonder whether the random field model with
$(N_c=2.83...$ in $d=4$ also gives rise to a Kosterlitz-Thouless
transition. Elsewhere,\cite{tissier06,tissier06b} we have answered
this question through an analysis of the perturbative FRG equations
near $d=4$ at $2$ loops, finding that there is no
Kosterlitz-Thouless-like transition. A practical consequence of this
result is that, as indeed reproduced by the numerical solution of the
truncated NP-FRG equations, the critical line $N_{lc}(d)$ drops
abruptly with an infinite slope as one moves away from $d=4$ (see
Figure~\ref{fig_diag_n_d}).

Going back now to the first question raised above, we conclude from
looking at Figure~\ref{fig_diag_n_d} that the $N=2$ RF$XY$M in $d=3$
is very distinctly in the region where no QLRO phase
occurs. Therefore, we find that there is no Bragg glass in this
model. This is confirmed by a direct study of our equations in $N=2,
d=3$.\cite{tissier06} As a cautionary note here, we would like to point
out that the absence of a Bragg glass phase in the $3-d$ RF$XY$M does
not necessarily implies that no such phase exists in disordered
type-$II$ superconductors. The relation between vortices in the latter
systems and the RF$XY$M is derived via an ``elastic glass
model'',\cite{giamarchi95} and it does not guarantee that phase
transitions associated with the presence of massive modes are
identical in the two systems.

\subsection{Robustness of the results}

The results presented above rely on a nonperturbative, but
approximate, RG description. It is therefore desirable to have some
estimate of its degree of accuracy. In addressing this point, we make
use of two main properties of the present NP-FRG formalism, which we
have emphasized in several occasions in this article and in paper I:

1) \textit{the existence of a systematic truncation scheme} that
allows one to control and improve the results by going to higher
orders of the truncation; this has been tested with success on the
pure $O(N)$ model\cite{berges02,canet03a, canet03b}, and we plan to
do in the near future for the RF$O(N)$M (the calculations being,
however, much more demanding).

2) \textit{a unified description of the whole $(N,d)$ plane} for the
$d$-dimensional RF$O(N)$M; one can therefore check the consistency of
the approximate nonperturbative results by comparing with known exact
or perturbative results in the appropriate regions of the $(N,d)$
plane. This is what we address in the following.

The numerical resolution of the truncated NP-FRG equations confirm the
property already stressed above and in paper I, that one recovers the
$1$-loop perturbative predictions near $d=6$, when $N\rightarrow
\infty$, and, more significantly, near $d=4$. A study of the $2$-loop
perturbative FRG equations\cite{ledoussal06,tissier06,tissier06b} near
$d=4$ unambiguously supports the scenario found here concerning the
presence of two nontrivial critical lines, one associated with the
breaking of dimensional reduction, $N_{DR}(d)$, starting downward from
$N=18$ when $d\rightarrow 4^+$, and one giving the lower critical
dimension for QLRO below $d=4$, $N_{lc}(d)$, arriving in $N_c=2.83...$
with an infinite slope when $d\rightarrow 4^-$. There are some
quantitative differences (see Ref.~[\onlinecite{tissier06b}]), but the
overall picture near $d=4$ is well captured by the present truncation.

To estimate the error made in locating the critical line $N_{DR}(d)$,
we have also considered a somewhat cruder approximation for the
$2$-replica potential $v_k$, which is a plain expansion in powers of
the fields (still around the minimum) including all terms up to
$\varphi^4$:
\begin{equation}
\begin{split}
v_k=2 v_{1,k}&(\sqrt{\rho_1\rho_2}z-\rho_{m,k})+v_{2,k}(\rho_1 +\rho _2-2\rho_{m,k})^2 \\&+ v_{3,k}(\rho_1 -\rho_2)^2+v_{4,k} (\sqrt{\rho_1 \rho_2}z-\rho_{m,k})^2\\&+ v_{5,k}(\sqrt{\rho_1
  \rho_2}z-\rho_{m,k})(\rho_1 +\rho _2-2\rho_{m,k}),
\end{split}
\end{equation}
where $v_{1,k}=1$ by construction. The results (obtained from
monitoring the divergence of $\partial^2_{\vert \vect \varphi_1 -
  \vect \varphi_2\vert^2}v_k\vert_{\vect \varphi_1 = \vect
  \varphi_2}$, see above) are very similar to those obtained with the
more involved truncation: the critical line $N_{DR}(d)$ starts again
from $N=18$ near $d=4$ and extends down to $d_{DR}\simeq 5$ when
$N=1$. The value of $d_{DR}(N=1)$ is found in the window $4.9-5.1$,
depending on the approximation and the choice of cutoff
function.\footnote{In addition, one should keep in mind that the
  dimensional-reduction fixed point is only approximately described in
  the minimal truncation (see paper I) and that the exponents $\eta$
  and $\bar \eta$ slightly deviate from the dimensional-reduction
  prediction as one moves away from $d=4$ and from $d=6$.}

Finally, one may also compare the lower critical value $N_{lc}(d)$
obtained when $d=2$ to the expected exact result, $N_{lc}(d=2)=1$. As
stated above, we find $N_{lc}(d=2)\simeq 1.15$, which provides an
estimate of the error. (The lower critical dimension of the RFIM is a
difficult an unfavorable test for the present approach starting from a
Ginzburg-Landau-Wilson bare action; as shown for the pure Ising model,
the results can be improved by solving the full first order of the
derivative expansion.\cite{ballhausen04})

The above discussion indicates that there is certainly room for
improvement of the quantitative predictions (and we have provided a
formalism to do so), but it also gives strong confidence in the
robustness of the present description of the long-distance physics of
the RF$O(N)$M.

\section{Physics of the cusp and role of temperature}
\label{sec_cusp}

We devote this section to a discussion of the physics associated with
the nonanalyticity found in the effective average action and of the
role of temperature. To proceed, we build upon the body of work
already done in the context of disordered elastic systems
\cite{balents93,balents96,chauve00,balents04,balents05}. For ease of notation, we focus on the
RFIM in the following, but most results will apply \textit{mutatis
  mutandis} to the RF$O(N)$M.

\subsection{Interpreting the cusp}

Breakdown of dimensional reduction has been associated with the
presence of multiple ``metastable states'', a metastable state being
generically taken as a field configuration that minimizes some action
or effective action. From the supersymmetric formalism of the
RFIM,\cite{ parisi79, parisi84b} one finds that a necessary condition
for this breakdown is the existence of many \textit{minima of the bare
  action} (as given by Eq. (1) of paper I). In the present approach on
the other hand, the failure of dimensional reduction predictions
originates in a strong enough nonanalyticity in the field dependence
of the dimensionless effective average action. To shed light on the
connection between this nonanaliticity and a picture in terms of
metastable states, we follow the line of reasoning developed for
elastic systems pinned by a random
potential.\cite{balents93,balents96,chauve00,balents04,balents05}

To begin with, it is worth stressing the unusual character of the RG
analysis in the presence of a cusp. Generally speaking, integration
over fluctuations, \textit{e.g.}, thermal fluctuations in Statistical
Physics, smooth away nonanalyticities as well as the effect of
possible metastable states so that, at long distance, the
dimensionless effective average action is a nonsingular function of
the fields. The novelty in the RFIM case comes from the dominance of
the (quenched) disorder fluctuations over the thermal ones and the
associated property that the long-distance physics (criticality,
ordering and quasi-ordering) is controlled by zero-temperature fixed
points. Actually, this physics is describable by working directly at
zero temperature at all scales: see above and paper I. As argued in
the context of disordered elastic systems,\cite{fisher86,balents93,balents96}
integration over high-energy modes along the RG flow amounts at zero
temperature to minimizing some coarse-grained action, and it is this
minimization procedure that may lead to cusp-like behavior in the
presence of multiple minima in the coarse-grained action.

It is now instructive to go back to the interpretation of the
$2$-replica potential $V_k(\phi_1,\phi_2)$ as the second cumulant of
the renormalized disorder and of its second derivative
$\Delta_k(\phi_1,\phi_2)$ as the second cumulant of the renormalized
random field, both being evaluated for uniform field configurations
(see paper I). To be more precise, the $1$-replica component of the
effective average action $\Gamma_{k,1} [\phi]$ is the Legendre
transform of the first moment of the random free energy functional at
scale $k$, $W_k[J; h]$, namely,
\begin{equation}
  \Gamma_{k,1}[\phi ] = - \overline{W_k[ J; h]} +  \int_{\vect x}  J(\vect x)\cdot  \phi(\vect x),
\end{equation}
where $J(\vect x)$ is a linear source conjugate to the field
$\phi(\vect x)$ and the overbar denotes an average over quenched
disorder, with $h$ denoting the bare random field. On the other hand,
the $2$-replica component is the second cumulant of $W_k[ J; h]$ with
$ J\equiv J_k[\phi]$, where $ J_k[\phi]$ is the nonrandom source
defined via the above Legendre transform, \textit{i.e.}, $
J_k[\phi](\vect x)=\delta \Gamma_{k,1} [\phi] / \delta \phi(\vect
x)$. One therefore has
\begin{equation}
  \label{eq_cumg2}
\Gamma_{k,2}[ \phi_1,  \phi_2] = \overline{\delta W_k[ J_k[\phi_1]; h]\delta W_k[ J_k[\phi_2]; h]},
\end{equation}
with $\delta W_k[ J; h]=W_k[ J; h]-\overline{W_k[ J; h]}$.

One can also define a renormalized random field at the running scale
$k$, $ \breve{ h}_k[ \phi](\vect x)$, as
\begin{equation}
 \breve{h}_k[ \phi](\vect x)=- \frac{\delta}{\delta \phi(\vect x)}\delta W_k[ J[\phi]; h].
\end{equation}
It has zero mean and its second cumulant is given by
\begin{equation}
  \label{eq_cumhren2}
\overline{\breve{h}_k[ \phi_1](\vect x) \breve{h}_k[ \phi_2](\vect y)}= \Gamma_{k,2,\vect x \vect y}^{(11)}[ \phi_1, \phi_2].
\end{equation}
More details, including a discussion of higher-order cumulants, can be
found in paper I.

In the truncated NP-FRG considered here, the random free energy
functional $\delta W_k[ J[\phi]; h]$ is taken in a local approximation
which amounts to replacing it by a random potential $\breve{V}_k(\phi;
\vect x)$ with zero mean and second cumulant
\begin{equation}
  \label{eq_cum_randompot_local}
\overline{\breve{V}_k[(\phi_1; \vect x) \breve{V}_k( \phi_2; \vect y)}\simeq \delta(\vect x- \vect y) V_k( \phi_1, \phi_2).
\end{equation}
Similarly, in this approximation, the renormalized random field
defined above is given by $\breve{h}_k(\phi;\vect x)= - \partial_\phi
\breve{V}_k[(\phi_1; \vect x)$ with a second cumulant
\begin{equation}
  \label{eq_cumhren2_local}
\overline{\breve{h}_k(\phi_1; \vect x) \breve{h}_k(\phi_2; \vect y)}\simeq \delta(\vect x- \vect y) \Delta_k( \phi_1, \phi_2).
\end{equation}

All the above considerations of course apply to the dimensionless
quantities, $v_k(\varphi_1,\varphi_2)$ and
$\delta_k(\varphi_1,\varphi_2)$. In particular, one can introduce a
dimensionless random potential $\breve{v}_k$ with its second cumulant
given by $v_k( \varphi_1, \varphi_2)$ and an associated dimensionless
random field with second cumulant given by $\delta_k( \varphi_1,
\varphi_2)$. In what follows, we rather discuss the dimensionless
functions since, at the fixed point, the dimensionful quantity $V_k$
goes to zero whereas $\Delta_k$ diverges.

\begin{figure}[htbp]
  \centering
  \subfigure[][]{\includegraphics[width=\linewidth]{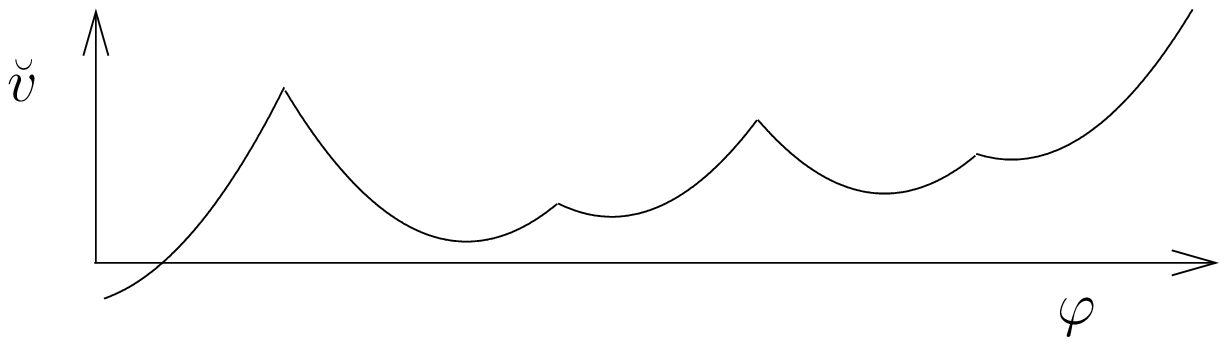}}
  \subfigure[][]{\includegraphics[width=\linewidth]{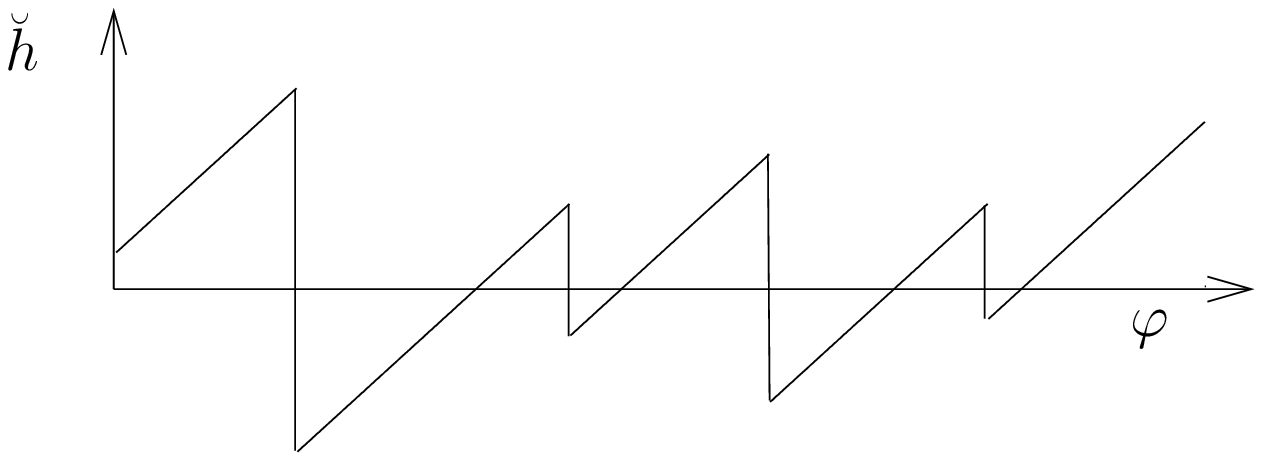}}
\caption{Sketch of the field dependence of the dimensionless
renormalized random potential $\breve{v}_*(\varphi)$ (a) and random
field $\breve{h}_*(\varphi)=-\partial_\varphi\breve{v}_*(\varphi)$ (b)
associated with a cusp in $\delta_*( \varphi_1, \varphi_2)$ as
$\varphi_2\rightarrow \varphi_1$. The cusps separating the minima in
(a) correspond to dicontinuities or ``shocks'' in (b); they are
located at random positions along the field axis.
}
\end{figure}
Following Ref.~[\onlinecite{balents96}], a cusp in $\delta_*( \varphi_1,
\varphi_2)$ as $\varphi_2\rightarrow \varphi_1$ can be interpreted as
resulting from a \textit{cuspy random potential}. Such a potential is
sketched in Figure 7a. It displays a sequence of minima separated by
cusps located at random positions along the field axis. (The explicit
spatial dependence of $\breve{v}_*(\varphi; \vect x)$ has been dropped
as for instance could be obtained from a properly rescaled integration
over space.) As seen in Figure 7b, the dimensionless random field then
shows discontinuities at those random locations., discontinuitites
that can be tnough of as ``shocks'' through an analogy with the
Burgers equation.\cite{balents96,ledoussal06b}

To see how the rugged cuspy landscape of Figure 7a gives rise to a
cusp in the second cumulant of the random field $\delta_*( \varphi_1,
\varphi_2)$, consider the quantity $\overline{(\breve{h}_*(\varphi_1)-
  \breve{h}_*(\varphi_2))^2}$ as $\varphi_2\rightarrow
\varphi_1$. Switching again to the variables
$x=(\varphi_1+\varphi_2)/2$ and $y=(\varphi_1-\varphi_2)/2$, one has
from Eq. (\ref{eq_cumhren2_local}):
\begin{equation}
\begin{split}
  \label{eq_cum_cusp}
\overline{(\breve{h}_*(\varphi_1)- \breve{h}_*(\varphi_2))^2}&\simeq 2(\delta_*( x,0)- \delta_*( x,y))\\& \simeq - 2 \delta_{*,a}( x) \vert y\vert + O(y^2),
\end{split}
\end{equation}
where we have used the cusp-like behavior as $y\rightarrow 0$ (see section III-B) and where $\delta_{*,a}( x)$ should be negative.

On the other hand, with a renormalized random field as pictured in
Figure 7b, $(\breve{h}_*(\varphi_1)- \breve{h}_*(\varphi_2))^2$ is a
$O(y^2)$ except when a discontinuity (shock) is present between the
two fields $\varphi_1$ and $\varphi_2$. As a consequence,
\begin{equation}
\begin{split}
  \label{eq_cum_shocks}
\overline{(\breve{h}_*(\varphi_1)- \breve{h}_*(\varphi_2))^2}\simeq \int_{x-\vert y\vert}^{x+\vert y\vert} dx_d \int d &\gamma_d \; \overline{p(x_d,\gamma_d)} \gamma_d^2 \\&+ O(y^2),
\end{split}
\end{equation}
where $x_d$ is the shock location and $\gamma_d$ the amplitude of the
associated discontinuity. Assuming that $\int d\gamma_d
\overline{p(x,\gamma_d)} \gamma_d^2$ is nonzero, one indeed recovers
Eq. (\ref{eq_cum_cusp}) with $\delta_{*,a}( x)=- \int d\gamma_d
\overline{p(x,\gamma_d)} \gamma_d^2$.

The above discussion therefore points to a picture in which, when
dimensional reduction is broken, the fixed point controlling the
critical behavior of the RFIM is described (for uniform field
configurations) by a dimensionless random potential with multiple
minima separated by cusps. These minima and the cuspy barriers
separating them arrive along the RG flow, because no minima are
present in the random potential at the microscopic scale $\Lambda$. It
is important to stress that this random potential is superimposed on a
mean dimensionless potential $u_*(\varphi)$ which itself displays two
equivalent minima at the fixed point. These global minima, located at
the dimensionless fields $\pm \varphi_{m,*}$ (see above), characterize
the incipient ferromagnetic ordering. Contrary to the disordered
elastic systems for which the random potential dominates at large
scales over the elastic energy, the random potential in random field
systems do controls the (sample-to-sample) fluctuations, but never
becomes predominant in the thermodynamic (mean) behavior. The
long-distance behavior is determined by a subtle interplay between
ferromagnetic ordering and randomness.

Finally, we mention that the rather elusive nature of the
nonanalyticity in the effective average action may become more
transparent when studied within a dynamic formalism. As pointed out by
Chauve et al.,\cite{chauve00} the presence of a cusp may then be
related to the existence of a nonzero threshold force below which the
system stays trapped in a minimum of the random potential
(``metastable state''). Again, this picture applies a zero temperature
and it describes an out-of-equilibrium dynamic transition, the
depinning of an elastic system in a random
environment.\cite{fisher85b} One may conjecture that a similar
relation exists in the RFIM between cusp-like behavior and
zero-temperature driven dynamics among metastable states: the analog
of the depinning transition would then be an ``avalanche transition''
observed when driving the system by slowly varying an applied magnetic
field (to sue the terminology of magnetic materials in which the
phenomenon is commonly observed).\cite{sethna06}

\subsection{Role of temperature, droplet phenomenology and activated dynamic scaling}

Temperature, as has been stressed in several occasions in this article
and in paper I, plays a peculiar role in random field systems. It is
irrelevant near the fixed points controlling criticality, ordering and
quasi-ordering. However, it is ``dangerously irrelevant'' in that, at
nonzero temperature and slightly off the critical point, both static
and dynamic quantities display somewhat anomalous scaling behavior
coming from the {\bf scale} dependence on the renormalized
temperature.\cite{villain85,fisher86b} To make contact with the
discussion of the preceding subsection, one may summarize the
situation as follows:\cite{balents04,balents05} the zero-temperature
analysis provides information on the typical behavior of the system
(including the typical fluctuations and correlation functions), as
described by its ground state, whereas small nonzero temperature
requires an account of rare events such as low-energy
excitations. Metastable states play a role in both cases: rather
intricate as far as the typical behavior is concerned (see above),
more direct in the case of low-energy excitations. In the latter case,
an efficient phenomenological approach has been proposed, known as the
``droplet picture''.\cite{bray84,fisher88}

In a nutshell, the droplet approach assumes the existence of rare
samples (or rare regions in a sample) for which, on top of the ground
state, an additional minimum (metastable state) is thermally
accessible, having an energy above the ground state of the order of
the temperature. If one defines for a system of linear size $L$
sample-dependent (\textit{i.e.}, random) ``connected'',
$\breve{\chi}_c[h]$, and ``disconnected'', $\breve{\chi}_d[h]$,
susceptibilities as
\begin{equation}
\label{eq_suscept_conn}
\breve{\chi}_c[h]=L^{-d}\int_{\vect x}\int_{\vect y}\bigg[\langle\chi(\vect x) \chi(\vect y)\rangle-
\langle\chi(\vect x)\rangle \langle\chi(\vect y)\rangle \bigg],
\end{equation}
and
\begin{equation}
\label{eq_suscept_disc}
\breve{\chi}_d[h]=L^{-d}\int_{\vect x}\int_{\vect y}
\langle\chi(\vect x)\rangle \langle\chi(\vect y)\rangle,
\end{equation}
where $\chi(\vect x)$ is the fundamental field in the bare action and
the brackets denote a thermal average for a given configuration $h$ of
the bare random field (see section II-A of paper I). At criticality
and at a temperature $T$ (criticality is attained by fine tuning the
bare disorder strength whose critical value depends on $T$ as
illustrated in Figure 1 of paper I), most samples, which are
characterized by a single populated minimum, are such that
$\breve{\chi}_d[h]\sim L^{4- \bar \eta}$ whereas, due to cancellation
of the leading terms in Eq. (\ref{eq_suscept_conn}),
$\breve{\chi}_c[h]\sim T L^{2- \eta}$. On the other hand, rare
samples, which occur with a propability assumed to be of the order of
$T L^{-\theta}$ with $\theta$ the temperature exponent given by
$\theta=2+\eta- \bar \eta$, have $\breve{\chi}_d[h]\sim
\breve{\chi}_c[h]\sim L^{4- \bar \eta}$ (since the leading
contributions to the two terms in $\breve{\chi}_c[h]$ no longer
cancel when two minima are populated). One therefore finds for the
disorder averaged $p$th moments:
\begin{equation}
\label{eq_scaling_suscept_conn}
\overline{\breve{\chi}_c^p}\sim T L^{p(4- \bar \eta)-\theta},
\end{equation}
\begin{equation}
\label{eq_scaling_suscept_disc}
\overline{\breve{\chi}_d^p}\sim L^{p(4- \bar \eta)}.
\end{equation}
As a consequence, the fluctuations of the connected susceptibility are
``anomalous'' with, \textit{e.g.}, $\overline{\breve{\chi}_c^2}\gg
\left(\overline{\breve{\chi}_c}\right)^2$.

Another important prediction of the droplet approach concerns the
dynamics of the RFIM near the critical point: the critical slowing
down of the relaxation is shown to be
``activated''.\cite{villain84,fisher86b,fisher88b} The typical
relaxation time diverges exponentially as one approaches the critical
point, with the effective activation barrier for relaxation diverging
with system size as\cite{fisher86b} $L^\theta$ at
criticality.\footnote{ This behavior is quite different than
  conventional critical slowing down for which the relaxation time
  diverges as a power law described by a critical exponent
  $z$.\cite{hohenberg77}}

A major step toward formulating a fully consistent field theory and
renormalization group framework for the droplet picture has been
accomplished by Balents and Le Doussal,\cite{balents04,balents04b,balents05,ledoussal06b} in the context of the random elastic
model. The core of the connection between FRG formalism and droplet
phenomenology is the existence of a ``thermal boundary layer'' that
governs the highly nonuniform limit of the renormalized temperature
going to zero. In the following, we do not attempt to provide an
exhaustive description of thermal boundary layer and droplet picture
in the RFIM, but rather stress some already illustrative results that
we obtain within the minimal NP-FRG truncation.

To study the role of temperature in the NP-FRG formalism, we consider
the flow equations for the RFIM, keeping now the terms depending on
the renormalized temperature. From the results of section IV-B of
paper I, one derives
\begin{equation}
  \label{eq_u'_ising_temp}
- \partial_t u_k' (x)= \beta_{u'}^{T=0}(x) + 2v_d T_k l_1^{(d)}(u''_k(x))u'''_k(x),
\end{equation}
\begin{equation}
 \label{eq_beta_delta_temp}
\begin{split}
- &\partial_t \delta_k(x,y) = \beta_{\delta}^{T=0}(x,y) + v_d T_k \bigg\{\frac{1}{2}[l_{1}^{(d)}(u''_+) +  l_{1}^{(d)}(u''_-)] \\&\times \delta_k^{(02)}(x,y) - [( l_{2}^{(d)}(u''_+) - l_{2}^{(d)}(u''_-)]  \delta_k^{(01)}(x,y) + \\&[l_{1}^{(d)}(u''_+) -  l_{1}^{(d)}(u''_-)] \delta_k^{(11)}(x,y)] + \frac{1}{2}[l_{1}^{(d)}(u''_+) + l_{1}^{(d)}(u''_-)]  \\& \times \delta_k^{(20)}(x,y)]-[( l_{2}^{(d)}(u''_+) + l_{2}^{(d)}(u''_-)]  \delta_k^{(10)}(x,y)\bigg\},
\end{split}
\end{equation}
where $\beta_{u'}^{T=0}$ and $\beta_{\delta}^{T=0}$ are the $T=0$ beta
functionals given by the right-hand sides of Eqs. (\ref{eq_u'_ising})
and (\ref{eq_delta_ising}) respectively in which the running
exponents $\eta_k$ and $\bar\eta_k$ are now expressed at $T \neq 0$,
and where $u''_{\pm}\equiv u''_k(x\pm y)$ as in
Eq. (\ref{eq_delta_ising}).

Taking the limit $y\rightarrow 0$ in Eq. (\ref{eq_beta_delta_temp})
and allowing for cusp-like behavior, one finds for
$\delta_{k,0}(x)=\delta_{k}(x,y=0)$ that
\begin{equation}
 \label{eq_flow_delta0_temp}
\begin{split}
- \partial_t &\delta_{k,0}(x) = \beta_{\delta_{0}}\vert_{reg}(x) \\&- \frac{v_d}{2} l_2^{(d)}(u''_k(x))\; \partial_y^2(\delta_k(x,y)- \delta_{k,0}(x))^2\vert_{y=0}
\\&+ v_d T_k  l_{1}^{(d)}(u''_k(x))\; \partial_y^2\delta_k(x,y)\vert_{y=0}
\end{split}
\end{equation}
with
\begin{equation}
 \label{eq_beta_delta0_reg_temp}
\begin{split}
\beta_{\delta_{0}}\vert_{reg}(x)= \beta_{\delta_{0}}^{T=0}&\vert_{reg}(x) + v_d T_k  \bigg[l_{1}^{(d)}(u''_k(x))\delta_{k,0}''(x)  \\& -
2 l_{2}^{(d)}(u''_k(x))\delta_{k,0}'(x)u'''_k(x) \bigg].
\end{split}
\end{equation}
and $\beta_{\delta_{0}}^{T=0}\vert_{reg}$ is given by the right-hand
side of Eq. (\ref{eq_delta0_ising}).

If a cusp appears in the renormalized cumulant $\delta_k(x,y)$ near
the fixed point, \textit{i.e.}, $\delta_k(x,y)= \delta_{k,0}(x) +
\vert y \vert \delta_{k,a}(x) + O(y^2)$ as in
Eq. (\ref{eq_delta_cusp}), then the two ``anomalous'' terms appearing
in the right-hand side of Eq. (\ref{eq_flow_delta0_temp}) behave quite
differently: the term characteristic of the $T=0$ behavior,
$\partial_y^2(\delta_k(x,y)- \delta_{k,0}(x))^2\vert_{y=0}$, goes to
$2 \delta_{*,a}(x)^2$, whereas that proportional to $T_k$,
$\partial_y^2\delta_k(x,y)\vert_{y=0}$, blows up. For a fixed point to
be reached and the theory be renormalizable, the latter divergence
must be cancelled. The solution is that (i) there should be no cusp at
finite $T_k$ and (ii) convergence to the cuspy $T=0$ fixed-point
function is nonuniform in $y$ as $T_k\rightarrow 0$ and takes the form
of a boundary layer.

In the close vicinity of the fixed point, when both $y$ and $T_k$
approach zero, one anticipates the following behavior:
\begin{equation}
\label{boundarylayer_delta}
\delta_k(x,y)= \delta_{*,0}(x)  + T_k f(x,\frac{y}{T_k}) + O(T_k^2)
\end{equation}
where $\delta_{*,0}(x)$ is the ($T=0$) fixed point result for $y=0$
and $f(x, \tilde{y})$ is a scaling function which is even in $x$ and
$\tilde{y}$ and analytic around $\tilde{y}=0$; $O(T_k^2)$ denotes
terms of order at least $T_k^2$ at fixed scaling variable
$\tilde{y}$. With Eq. (\ref{boundarylayer_delta}), the two anomalous
contributions described above can be expressed as
\begin{equation}
\label{anomalousT=0}
\partial_y^2(\delta_k(x,y)- \delta_{k,0}(x))^2\vert_{y=0}= 2 f(x,0) f^{(02)}(x,0),
\end{equation}
\begin{equation}
\label{anomalousT_k}
T_k \partial_y^2\delta_k(x,y)\vert_{y=0}= f^{(02)}(x,0),
\end{equation}
with the same convention as before for the notation of partial
derivatives.

Inserting the above expressions into the flow equation for
$\delta_k(x,y)$, Eq. (\ref{eq_beta_delta_temp}), and using the fact
that $\delta_{k,0}(x)$ is solution of the the fixed-point equation at
zero temperature for $y=0$, one finds up to a $O(T_k)$ that the
scaling function $f$ must satisfy that
\begin{equation}
\begin{split}
\label{eq_scalingfunction}
\frac{1}{2} l_{2}^{(d)}(u''_*(x))\;&\partial_{\tilde{y}}^2( f(x,\tilde{y})-f(x,0))^2 \\&- l_{1}^{(d)}(u''_*(x)) f^{(02)}(x,\tilde{y})
\end{split}
\end{equation}
is independent of $\tilde{y}$.

The solution is easily obtained as
\begin{equation}
\begin{split}
\label{scalingfunction}
&f(x,\tilde{y})-f(x,0)= \\&\frac{l_{1}^{(d)}(u''_*(x))}{l_{2}^{(d)}(u''_*(x))}\left[ 1-\sqrt{1-\left( \frac{l_{2}^{(d)}(u''_*(x))f^{(02)}(x,0))}{l_{1}^{(d)}(u''_*(x))}\right) \tilde{y}^2 }\right],
\end{split}
\end{equation}
with $f^{(02)}(x,0)<0$ and $u''_*(x)$ given by the fixed-point
solution of Eq. (\ref{eq_u'_ising_temp}). The zero-temperature cuspy
fixed point is recovered by considering $y\neq 0$ and $T_k\rightarrow
0$, which leads to
\begin{equation}
\label{scalingfunction_asympt}
 f(x,\tilde{y}\rightarrow \pm \infty) \sim  \delta_{*,a}(x) \vert \tilde{y}\vert
\end{equation}
with $\delta_{*,a}(x)= f^{(02)}(x,0) < 0$. On the other hand, when
$y\rightarrow 0$ at fixed $T_k\neq 0$, \textit{i.e.}, near
$\tilde{y}=0$,
$f(x,\tilde{y})-f(x,0)=O(\tilde{y}^2)$. Eq. (\ref{scalingfunction})
thus describes the rounding of the cusp near $y=0$ in a layer whose
width of order $T_k$ goes to zero as $k\rightarrow 0$.

We can now make contact with the droplet description of the RFIM. From
the effective average action $\Gamma_k$, one has access to all Green
functions of the system via the $1-PI$ vertices (see paper I). In the
present minimal truncation,
\begin{equation}
\label{eq_truncation_Gamma}
\begin{split}
  &\Gamma_k\left[\{\phi_a\}\right ]= \int_{\vect x}
  \bigg\{\frac{1}{T}\sum_{a=1}^n\big[ \frac 1{2} Z_{m,k}
  \vert \partial \phi_a(\vect x)\vert ^2 + U_{k}(\phi_a(\vect x))
  \big]\\& -\frac 1{2T^2} \sum_{a,b=1}^nV_{k}( \phi_a(\vect x),
  \phi_b(\vect x)) \bigg\},
\end{split}
\end{equation}
where we use an explicit dependence on a bare temperature $T$ for
book-keeping purpose.

The $1-PI$ vertices are obtained by functional differentiation and
their expression is given in Appendix B. When all fields are taken as
uniform and, moreover, equal, the first vertices have the following
form:
\begin{equation}
\label{eq_unif_Gamma2}
\begin{split}
\Gamma_{k,(a,\vect q_1)(b, \vect q_2)}^{(2)}&\left(\{\phi_f=\phi\}\right )=(2\pi)^d \delta(\vect q_1+\vect q_2)  \\& \bigg\{\delta_{ab}\; \widehat{\Gamma}_k^{(2)}(\phi; q_1)+ \widetilde{\Gamma}_k^{(2)}(\phi, \phi;q_1) \bigg\},
\end{split}
\end{equation}
\begin{equation}
\label{eq_unif_Gamma3}
\begin{split}
\Gamma_{k,(a,\vect q_1)(b, \vect q_2)(c,\vect q_3)}^{(3)}&(\{\phi_f=\phi\})= \\& \;\;(2\pi)^d \delta(\vect q_1+\vect q_2+\vect q_3) \Gamma_{k,abc}^{(3)}(\phi),
\end{split}
\end{equation}
\begin{equation}
\label{eq_unif_Gamma4}
\begin{split}
\Gamma^{(4)}&_{k,(a,\vect q_1)(b, \vect q_2)(c,\vect q_3)(d,\vect q_4)}(\{\phi_f=\phi\} )=\\& \;\;\;\;\; \;\;\;\;\;\;\;\;\;(2\pi)^d \delta(\vect q_1+\vect q_2+\vect q_3+\vect q_4)\Gamma_{k,abcd}^{(4)}(\phi),
\end{split}
\end{equation}
where the functions $\widehat{\Gamma}_k^{(2)}(\phi;q)$,
$\widetilde{\Gamma}_k^{(2)}(\phi, \phi;q)$,
$\Gamma_{k,abc}^{(3)}(\phi)$, $\Gamma_{k,abcd}^{(4)}(\phi)$ can be
derived from Eqs.~(\ref{eq_unif_Gamma2ab}--\ref{eq_truncation_Gamma4abcd}).

The (connected) Green functions may be obtained from the replicated
free energy functional $W_k[\{J_a\}]$, which is the Legendre transform
of the effective average action $\Gamma_{k}[\{\phi_a \}]$. The
procedure consists of using the standard formulas that relate the
$W_k^{(p)}$'s to the $\Gamma_{k}^{(p)}$'s\cite{zinnjustin89} and
expanding both sides in number of free replica sums as explained in
paper I, keeping only the leading terms. This is detailed in Appendix
B. The Green functions can be cast in a form similar to that of the
$1-PI$ vertices, namely, for equal field arguments,
\begin{equation}
\label{eq_truncation_W2}
\begin{split}
  W&_{k,(a,\vect q_1)(b, \vect q_2)}^{(2)}(\{\phi_f=\phi\})=(2\pi)^d
  \delta(\vect q_1+\vect q_2) \times \\& \bigg\{ \delta_{ab}\;
  \widehat{G}_k^{(2)}(\phi; q_1)+ \widetilde{G}_k^{(2)}(\phi,
  \phi;q_1) + O(\sum_{f}) \bigg\},
\end{split}
\end{equation}
\begin{equation}
\label{eq_unif_W3}
\begin{split}
W_{k,(a,\vect q_1)(b, \vect q_2)(c,\vect q_3)}^{(3)}&(\{\phi_f=\phi\})= \\& (2\pi)^d \delta(\vect q_1+\vect q_2+\vect q_3) G_{k,abc}^{(3)}(\phi;\vect q_1,\vect q_2),
\end{split}
\end{equation}
\begin{equation}
\label{eq_unif_W4}
\begin{split}
W&_{k,(a,\vect q_1)(b, \vect q_2)(c,\vect q_3)(d,\vect q_4)}^{(4)}(\{\phi_f=\phi\} )=\\& \;\;\; (2\pi)^d \delta(\vect q_1+\vect q_2+\vect q_3+\vect q_4)G_{k,abcd}^{(4)}(\phi;\vect q_1,\vect q_2,\vect q_3),
\end{split}
\end{equation}
where the various quantities appearing in the right-hand sides are
related to the $1-PI$ counterparts in
Eqs. (\ref{eq_unif_Gamma2}-\ref{eq_unif_Gamma4}) as discussed in Appendix B. 
Note that in the present truncation of the effective average
action (limited to the first order of the derivative
expansion),\cite{tarjus07_1} information on the Green functions is
essentially limited to zero external momenta, more precisely external
momenta with $\vert \vect q\vert$ less than the running scale
$k$. Thanks to the RG framework, this is enough to provide a
determination of the anomalous dimension of the field $\eta$, but in
what follows, we only consider the case of external momenta set to
zero.

Next, focusing on the critical (scaling) region, we introduce
dimensionless functions and fields by using the scaling dimensions
suitable for a zero-temperature fixed point (see section II); so, for
instance,
\begin{equation}
\label{eq_scalingU''}
\frac{1}{T}U_k''(\phi)\simeq \frac{k^{4- \bar\eta}}{T_k}u_*''(\varphi),
\end{equation}
\begin{equation}
\label{eq_scalingDelta}
\frac{1}{T^2} \Delta_{k}(\phi_1, \phi_2)\simeq \frac{k^{4- \bar\eta}}{T_k^2}\delta_{k}(\varphi_1, \varphi_2).
\end{equation}
In addition, we eventually take the limit of equal field arguments in
all expressions after insertion of the results obtained for the
thermal boundary layer description of $\delta_{k}(\varphi_1,
\varphi_2)$, Eqs. (\ref{boundarylayer_delta},\ref{scalingfunction}).

In the scaling region, the Green functions at zero external momenta
can then be expressed as
\begin{equation}
\label{eq_scaled_hatG2}
\widehat{G}_k^{(2)}(\phi,0)\simeq T_k k^{-(4-\bar\eta)} h_{*}^{(2)}(\varphi),
\end{equation}
\begin{equation}
\label{eq_scaled_tildeG2}
\widetilde{G}_k^{(2)}(\phi, \phi;0)\simeq k^{-(4-\bar\eta)} g_{*}^{(2)}(\varphi),
\end{equation}
\begin{equation}
\label{eq_scaled_G3}
\begin{split}
G_{k,abc}^{(3)}(\phi; \vect 0,\vect 0)\simeq k^{-\frac{d}{2}-\frac{3}{2}(4-\bar\eta)}&\bigg\{g_{*}^{(3)}(\varphi)+T_k h_{*}^{(3)}(\varphi)(\delta_{ab}\\&+ \delta_{bc}+ \delta_{ca}) + O(T_k^2) \bigg\},
\end{split}
\end{equation}
\begin{equation}
\label{eq_scaled_G4}
\begin{split}
G&_{k,abcd}^{(4)}(\phi; \vect 0,\vect 0,\vect 0)\simeq k^{-d-2(4-\bar\eta)}\bigg\{g_{*}^{(4)}(\varphi)+T_k h_{*}^{(4)}(\varphi)\times \\&(\delta_{ab}+ \delta_{ac}+ \delta_{ad}+ \delta_{bc}+ \delta_{bd}+ \delta_{cd}) + T_k \frac {f^{(02)}(\varphi,0)}{u_{*}''(\varphi)^4}\\&+ O(T_k^2) \bigg\},
\end{split}
\end{equation}
where $\phi \rightarrow 0$ (as $k^{(d-4+\bar\eta)/2)}\varphi$) and
where the fonctions $g_{*}^{(p)}(\varphi)$, $h_{*}^{(p)}(\varphi)$,
$p=2,3,4,...$ are obtained from $u_*''(\varphi)$,
$\delta_{*,0}(\varphi)$ and their derivatives. Their expression is not
particularly illuminating and we do not reproduce them here; see
Appendix B for more details.

From the definition of the replicated generating functional $W_k[\{J_a
\}]$, one can derive the relation between the replica Green functions
considered above and the physical Green functions directly defined in
the disordered system.\cite{tarjus07_1} Still working at the running scale
$k$ and at zero external momenta, the first moments of the (random)
``connected'' and ``disconnected'' susceptibilities introduced in
Eqs. (\ref{eq_suscept_conn},\ref{eq_suscept_disc}) are for instance
given by
\begin{equation}
  \label{eq_suscep_conn_moment1}
\overline{\breve{\chi}_{k,c}}=\widehat{G}_k^{(2)}(\phi;0),
\end{equation}
\begin{equation}
\label{eq_suscep_disc_moment1}
\overline{\breve{\chi}_{k,d}}=\widetilde{G}_k^{(2)}(\phi, \phi;0),
\end{equation}
whereas the second moments read
\begin{equation}
\label{eq_suscep_conn_moment2}
\begin{split}
&\overline{\breve{\chi}_{k,c}^2}= k^d \big[G_{k,aabb}^{(4)}(\phi)-2 G_{k,aabc}^{(4)}(\phi)+G_{k,abcd}^{(4)}(\phi)\big],
\end{split}
\end{equation}
\begin{equation}
\label{eq_suscep_disc_moment2}
\overline{\breve{\chi}_{k,d}^2}= k^d G_{k,abcd}^{(4)}(\phi),
\end{equation}
where distinct replica indices here mean distinct replicas (no
summation implied).

Putting together all the above results,
Eqs. (\ref{eq_scaled_hatG2}-\ref{eq_suscep_disc_moment2}), we find
that the moments of the random ``disconnected'' suceptibility scale as
\begin{equation}
\label{eq_suscept_disc_scaled1}
\overline{\breve{\chi}_{k,d}}\sim  k^{-(4- \bar \eta)},
\end{equation}
\begin{equation}
\label{eq_suscept_disc_scaled2}
\overline{\breve{\chi}_{k,d}^2}\sim  k^{-2(4- \bar \eta)},
\end{equation}
whereas those of the  random ``connected'' susceptibility scale as
\begin{equation}
\label{eq_suscept_conn_scaled1}
\overline{\breve{\chi}_{c,d}}\sim  T_k k^{-(4- \bar \eta)}\sim  T k^{-(2- \eta)},
\end{equation}
\begin{equation}
\label{eq_suscept_conn_scaled2}
\begin{split}
\overline{\breve{\chi}_{k,d}^2}&\sim  T_k k^{-2(4- \bar \eta)}\frac {f^{(02)}(\varphi,0)}{u_{*}''(\varphi)^4}\\& \sim T k^{-2(4- \bar \eta)+ \theta},
\end{split}
\end{equation}
where we have used that $T_k\sim T k^\theta$, and we recall that
$f^{(02)}(\varphi,0)=\delta_{*,a}''(\varphi) < 0$. Notice that only
the term due to the boundary layer appears in
Eq. (\ref{eq_suscept_conn_scaled2}) (the other contributions cancel
out).

From this analysis, we therefore obtain that the moments of the random
``connected'' and ``disconnected'' susceptibilities in the truncated
NP-FRG precisely scale as in the droplet description: compare
Eqs. (\ref{eq_suscept_disc_scaled1}-\ref{eq_suscept_conn_scaled2}) and
Eqs. (\ref{eq_scaling_suscept_conn},\ref{eq_scaling_suscept_disc}),
with $L\sim k^{-1}$. The ``anomalous'' scaling of the moments of the
``connected'' susceptibility, which is due to rare low-energy
excitations in the droplet picture, results in the NP-FRG from the
presence of a thermal boundary layer in the vicinity of the
zero-temperature fixed point, as illustrated by
Eq. (\ref{eq_suscept_conn_scaled2}). This is in complete agreement
with the more detailed analysis performed in
Ref.~[\onlinecite{balents05}] for disordered elastic systems.

To conclude this section, we briefly address the question of the
slowing down of the relaxation toward equilibrium near the critical
point. At long times, the dynamics of the RFIM can be modeled by a
Langevin equation,
\begin{equation}
\label{eq_langevin}
\partial_\tau \chi(\vect x, \tau)= - \Omega \frac{\delta S[\chi;h]}{\delta \chi(\vect x, \tau)} + \zeta(\vect x, \tau),
\end{equation}
where $\tau$ denotes the physical time (to be distinguished from the RG ``time'' $t$) and $\zeta(\vect x, \tau)$ is a thermal noise taken with a gaussian distribution characterized by a zero mean and a second moment
\begin{equation}
  \label{eq_cumulant_noise}
\overline{\zeta(\vect x,\tau)\zeta(\vect y,\tau')}=2 T \Omega \delta(\tau-\tau') \delta(\vect x-\vect y).
\end{equation}
In Eq. (\ref{eq_langevin}) $S[\chi;h]$ is the bare action for the
RFIM, with $h$ being the bare random field (see for instance section
II of paper I), and $\Omega$ is a kinetic coefficient that describes
the bare relaxation rate and sets the elementary time scale in the
problem. (We consider here the case of a nonconserved order
parameter,\cite{hohenberg77} but the case of a conserved order
parameter could also be of interest.\cite{huse87})

An RG formalism can be conveniently implemented by using standard
field theoretical techniques to build the generating functional of the
time-dependent correlation and response
functions.\cite{martin73,janssen76,dedominicis76,zinnjustin89} Associated with this functional is a ``bare dynamic action''
that depends on two fields, the fundamental field and a ``response''
field. The average over the quenched disorder can now be performed
without introducing replicas. By a Legendre transform, one then
introduces an ``effective dynamic action'' which is the generating
functional of the disorder-averaged, time-dependent $1-PI$ vertices.

In this setting, one can repeat the steps detailed in paper I to
construct a NP-FRG approach to the dynamics: add a mass-like regulator
with an infrared cutoff function that suppresses the contribution of
low-momentum and low frequency modes, define a (dynamic) effective
average action at scale $k$, whose evolution with $k$ is governed by
an exact RG flow equation, devise a truncation scheme.\cite{delamotte05} In the present problem, this latter step can be done
in the spirit of the minimal trucation considered above and in paper
I. One however needs an additional assumption concerning the time
dependence which, similarly to the spatial dependence, can be handled
via an appropriate ``derivative expansion''. The simplest
approximation that captures the physics is a ``single time scale''
approximation in which one introduces a single renormalized relaxation
rate $\Omega_k$. (This parallels the single wavefunction
renormalization parameter used to describe the spatial dependence of
the field in the minimal truncation: see paper I.)

From the boundary layer structure and by analogy with the previous
work on the random elastic model,\cite{chauve00,balents05} one then
expects that the renormalized relaxation rate flows as
\begin{equation}
\label{eq_flow_relaxation}
\partial_t ln(\Omega_k) \sim -T_k^{-1}
\end{equation}
near the zero-temperature fixed point. This indeed corresponds to
activated dynamic scaling and fits in with the droplet picture
summarized above. A proper derivation of this result and an account of
the (expected) broad distribution of relaxation rates would require a
detailed dynamic treatment, but this goes beyond the scope of the
present article.

\section{Conclusion and perspectives}
\label{sec_conclusion}

In this paper, which is the second part of a series of articles
reporting our work on a nonperturbative functional renormalization
group (NP-FRG) approach for random field models and related disordered
systems, we have applied the formalism presented in paper I to the
$d$-dimensional random field $O(N)$ model. We have focused on two main
issues related to the long-distance physics of the model: the
breakdown of the dimensional reduction property predicted by
conventional perturbation theory and the nature of the phase diagram
and ordering transitions in the $(N,d)$ plane.

Within our NP-FRG approach, the way out of dimensional reduction is
the appearance of a strong enough nonanalyticity in the field
dependence of the dimensionless effective average action near the
relevant zero-temperature fixed point. We have shown that this occurs
below a critical dimension $d_{DR}(N)$, which goes continuously from
$N_{DR}=18$ as $d\rightarrow 4^+$ to $d_{DR}\simeq 5$ for $N=1$.  In
addition, we provide a description of criticality, ferromagnetic
ordering, and quasi-long range order in the whole $(N,d)$ plane. The
NP-FRG method is able to directly address the phase diagram of the
model in low (physical) dimension $d$ and small (physical) number of
components $N$: in particular, we find that there is no
``Bragg-glass'' phase, \textit{i.e.}, no phase with quasi-long range
order in the $3$-dimensional RF$XY$M. Note that all those results are
made possible by the very structure of the present RG approach which
is (1) functional, (2) approximate but nonperturbative, and (3)
devised to provide a continuous and consistent description of the
whole plane of $N,d$.

Building upon earlier work on random elastic
models,\cite{balents04,balents05} we have also shown how the NP-FRG
formalism gives access to both the typical behavior of the system,
controlled by zero-temperature fixed points with a nonanalytic
dimensionless effective action, and to the physics of rare low-energy
excitations (``droplets''), described at nonzero temperature by the
rounding of the nonanalyticity in a thermal boundary layer.

Work still remains to be done for a complete understanding of random
field models. We have pointed in several occasions in this paper and
in the preceding one that clarifying the putative link between
breaking of the underlying supersymmetry\cite{parisi79} and appearance
of a nonanalyticity in the dimensionless effective action would
require to ``upgrade'' the present NP-FRG formalism to a superfield
formulation of the random field models. We defer this, as well as the
study of improved nonperturbative truncations, to a forthcoming
publication. We have also indicated an interesting extension of the
present work to the dynamics of the random field Ising model (RFIM),
both to the out-of equilibrium driven dynamics at zero temperature and
to the (activated) relaxation to equilibrium at nonzero
temperature. Finally, the connection to the proposed picture of the
RFIM in terms of ``spontaneous replica symmetry breaking'' and
``replica bound states''\cite{mezard92,dedominicis95,brezin01,parisi02} remain to be investigated.

The NP-FRG formalism appears as a powerful tool to study random field
models and related disordered systems. Whether such an approach can be
generalized to tackle another major unsettled problems of the field of
disordered systems, the long-distance physics of spin glasses, is a
challenging but completely open question.

We thank D. Mouhanna for helpful discussions.

\appendix
\section{Nonanalyticity in the RFIM near $d=6$}

Our starting point is the RG flow equations for $\delta_{k,0}(x)$ and
$\delta_{k,2p}(x)$ obtained by assuming that $\delta_k(x,y)$ is
regular enough near $y=0$: see
Eqs. (\ref{eq_delta0_ising}-\ref{eq_delta2p_ising}). The linear
operators appearing in those equations are given by
\begin{equation}
\begin{split}
\label{eq_linear2p_ising} 
&L_{2p}[u'',\delta_{0},\delta_{2}]= - \left[p(d-4+\bar \eta_k)+2
  \eta_k-\bar \eta_k \right]-\\&\frac{1}{2}(d-4+\bar\eta_k)
x \partial_x + 2v_d
\bigg\{l_{2}^{(d)}(u''(x))\delta_{0}(x)\partial_x^2 + 2(p+1) \\&
\times \bigg[l_{2}^{(d)}(u''(x))\delta_{0}'(x)-2l_{3}^{(d)}(u''(x))
\delta_{0}(x) u'''_k(x)\bigg]\partial_x + \\& \frac{p(2p+3)+1}{2}
\bigg[l_{2}^{(d)}(u''(x))\delta_{0}''(x)-2l_{3}^{(d)}(u''(x))
\delta_{0}'(x) \times \\& u'''_k(x)+ 2l_{4}^{(d)}(u''(x))
\delta_{0}(x) u'''_k(x)^2 \bigg] - \frac{p(2p+3)}{2}\times \\&
l_{2}^{(d)}(u''(x))\delta_{2}(x) \bigg\}
\end{split}
\end{equation}
for $p\geq 2$; the expression of $L_2[u'',\delta_0]$ is obtained by
setting $p=1$ in the above equation and dropping the last term so that
$\delta_2(x)$ no longer appears in the operator. Note that, as for the
threshold functions (see paper I), there is an explicit dependence on
$k$ due to $\eta_k$ and $\bar \eta_k$ that comes on top of the
dependence that may occur through the arguments.

Near $d=6$, one finds, as developed in section V-A of paper I, that
the fixed point is characterized by $\eta_*=\bar\eta_*=O(\epsilon^2)$,
$u_*''(x)=\epsilon(\lambda_{1*}/2)(3x^2-x_{m*}^2) + O(\epsilon^2)$,
$\delta_*(x,y)=1+\epsilon^2d(x,y)$, with $x_{m*}^2=6v_6l_2^{(6)}(0)$,
$\lambda_{1*}=(36 v_6 l_3^{(6)}(0))^{-1}$, and $d(x,y)=O(1)$. The
result for $\delta_*(x,y)$ implies that
$\delta_{*,0}'(x)=O(\epsilon^2)$ and $\delta_{*,2p}(x)=O(\epsilon^2)$.

After inserting these results in Eq. (\ref{eq_linear2p_ising}), one
obtains that for $p=O(1)$,
\begin{equation}
\label{eq_linear2p_O(1)} 
L_{2p*}(x) \simeq - 2p - x \partial_x + 2v_d  l_{2}^{(d)}(0) \partial_x^2+ O(\epsilon)
\end{equation}
at the fixed point. We have made the implicit assumption, whose
consistency can be checked, that the derivatives with respect to $x$
acting on the $\delta_{k,2p}$'s do not modify the order in
$\epsilon$. The eigenfunctions of the above linear operator (with the
condition that they are bounded by polynomials at large values of the
argument\cite{morris98b} are the Hermite polynomials $H_n(x/\sqrt{4v_d
  l_{2}^{(d)}(0)})$\cite{abramowitz64} with associated eigenvalues
$\lambda_{2p,n}=-(2p+n)$, with $n\in \nbN$. Recalling that the RG time
$t$ in Eq. (\ref{eq_delta2p_ising}) goes to $-\infty$ as $k\rightarrow
0$, the above result means that the corresponding directions are
irrelevant on approaching to the fixed point. (This confirms the
result found in section V-A of paper I that the fixed point given
above is once unstable at first order in $\epsilon$.)

However, a new phenomenon may appear when $p$ is very large and scales
as $1/\epsilon^2$. In this case, one finds that
\begin{equation}
\begin{split}
\label{eq_linear2p_large} 
L_{2p*}(x) \sim - 2p \bigg\{1 - \frac{v_d}{4} p \big[ &2
l_{4}^{(d)}(0) u_*'''(x)^2 + l_{2}^{(d)}(0) \times
\\&(\delta_{0*}''(x)- \delta_{2*}(x))\big] + O(\epsilon) \bigg\},
\end{split}
\end{equation}
where the whole second term in the braces is of order $1$ and, if
positive, can become larger than $1$ for some value of $p$ so that
$L_{2p*}$ becomes positive.

To analyze the sign of $L_{2p*}$, one has to study $\delta_{0*}''(x)$
and $\delta_{2*}(x)$ at order $\epsilon^2$. This is easily performed
from Eqs. (\ref{eq_delta0_ising}) and (\ref{eq_delta2_ising}). One
finds that $\delta_{0*}''(x)=-\delta_{2*}(x)=18 v_d
l_{4}^{(d)}(0)\lambda_{1*}^2 \epsilon^2$, which indeed guarantees that
the second term in the right-hand side of
Eq. (\ref{eq_linear2p_large}) is positive. Evaluating $L_{2p*}$ for
$x=x_{m*}$ and using the expression of $u_*'''(x)$ now gives
\begin{equation}
\label{eq_linear2p_large_min} 
L_{2p*}(x_{m*}) \sim - 2p \bigg\{1 -  K^2 (p\epsilon^2) + O(\epsilon) \bigg\},
\end{equation}
with $K=l_{2}^{(d)}(0) l_{4}^{(d)}(0)/ (6 l_{3}^{(d)}(0))^2$. One
therefore concludes that for $p\gtrsim 1/(K\epsilon)^2$
$L_{2p*}(x_{m*})$ becomes positive, which, according to
Eq. (\ref{eq_delta2p_ising}), leads to a divergence of
$\delta_{k,2p}(x_{m*})$ as $k\rightarrow 0$.  As a consequence, the
renormalized disorder cumulant displays a subcusp of order
$1/\epsilon^2$ at the fixed point. A related phenomenon has also been
proposed by Feldman.\cite{feldman02}

\section{Green functions in the truncated NP-FRG of the RFIM}

The $1-PI$ vertices are obtained by functional differentiation of the
truncated effective average action given in
Eq. (\ref{eq_truncation_Gamma}).  For uniform field configurations one
finds
\begin{equation}
\label{eq_unif_Gamma2ab}
\begin{split}
  \Gamma&_{k,(a,\vect q_1)(b, \vect q_2)}^{(2)}\left(\{\phi_f\}\right
  )=(2\pi)^d \delta(\vect q_1+\vect q_2) \times \\& \bigg\{\delta_{ab}
  \frac{1}{T}\big[ Z_{m,k} q_1^2 + U_{k}''(\phi_a) \big] -\frac 1{T^2}
  \Delta_{k}( \phi_a, \phi_b) \bigg\},
\end{split}
\end{equation}
\begin{equation}
\label{eq_truncation_Gamma3abc}
\begin{split}
  \Gamma&_{k,(a,\vect q_1)(b, \vect q_2)(c,\vect
    q_3)}^{(3)}(\{\phi_f\} )= (2\pi)^d \delta(\vect q_1+\vect
  q_2+\vect q_3)\times \\& \bigg\{\frac{\delta_{abc}}{T}
  U_{k}'''(\phi_a) -\frac {1}{2T^2} \big[\delta_{ab}
  \Delta_{k}^{(10)}(\phi_b, \phi_c) + perm(abc) \big] \bigg\},
\end{split}
\end{equation}
\begin{equation}
\label{eq_truncation_Gamma4abcd}
\begin{split}
  &\Gamma_{k,(a,\vect q_1)(b, \vect q_2)(c,\vect q_3)(d,\vect
    q_4)}^{(4)}\left(\{\phi_f\}\right )= (2\pi)^d \delta(\vect
  q_1+\vect q_2+\vect q_3+\vect q_4) \\&
  \bigg\{\frac{\delta_{abcd}}{T} U_{k}''''(\phi_a) -\frac 1{2T^2}
  \big[\delta_{abc}\Delta_{k}^{(20)}(\phi_c, \phi_d) + perm(abcd)
  \big]\\&-\frac 1{2T^2}
  \big[\delta_{ab}\delta_{cd}\Delta_{k}^{(11)}(\phi_a, \phi_c) +
  \delta_{ac}\delta_{bd}\Delta_{k}^{(11)}(\phi_a, \phi_d)\\&+
  \delta_{ad}\delta_{bc}\Delta_{k}^{(11)}(\phi_a, \phi_c)\big]
  \bigg\},
\end{split}
\end{equation}
where $\delta_{abc}\equiv \delta_{ab}\delta_{bc}$,
$\delta_{abcd}\equiv \delta_{ab}\delta_{bc}\delta_{cd}$,
``$perm(abc)$'' denotes the two terms obtained by circular
permutations of the indices $abc$, and ``$perm(abcd)$'' denotes the
three terms obtained by circular permutations of the indices
$abcd$. All other notations are as in paper I and above.

The connected Green functions $W_{k,(a_1,\vect q_1)...(a_p, \vect
  q_p)}^{(p)}(\{\phi_f\})$ are related to the $1-PI$ vertices by
formulas deriving from the Legendre transform between $W_k$ and
$\Gamma_k$.\cite{zinnjustin89} For instance, the $2$-point connected
Green function $W_{k}^{(2)}$ is the inverse of the $2$-point $1-PI$
vertex, $\vect W_k^{(2)}=\vect \Gamma_k^{(2)-1}$. By using the
expansion in number of free replica sums detailed in section II-D of
paper I, one obtains at leading order
\begin{equation}
\label{eq_truncation_W2ab}
\begin{split}
  W_{k,(a,\vect q_1)(b, \vect q_2)}^{(2)}(\{\phi_f\})=&(2\pi)^d
  \delta(\vect q_1+\vect q_2) \bigg\{ \delta_{ab}
  \widehat{G}_k^{(2)}(\phi_a; q_1)\\&+ \widetilde{G}_k^{(2)}(\phi_a,
  \phi_b;q_1) + O(\sum_{f}) \bigg\},
\end{split}
\end{equation}
where $O(\sum_{f})$ denotes higher orders in the expansion in number
of free replica sums and with
\begin{equation}
\label{eq_truncation_hatW2}
\widehat{G}_k^{(2)}(\phi_a ;q)=\frac{T}{ Z_{m,k} q^2 + U_{k}''(\phi_a)},
\end{equation}
\begin{equation}
\label{eq_truncation_tildeW2}
\widetilde{G}_k^{(2)}(\phi_a, \phi_b ;q)=\frac{\Delta_{k}(\phi_a, \phi_b)}{( Z_{m,k} q^2 + U_{k}''(\phi_a))( Z_{m,k} q^2 + U_{k}''(\phi_b))}.
\end{equation}
The $3$- and $4$-point connected Green functions are derived along
similar lines, using the standard graphical
representation\cite{zinnjustin89} and keeping the lowest order in the
expansion in free replica sums.

Consider now the scaling region. As discussed in the main text around
Eqs. (\ref{eq_scalingU''}, \ref{eq_scalingDelta}), one can introduce
dimensionless quantities and use the results concerning the thermal
boundary layer. For the $1-PI$ vertices evaluated at zero external
momenta and for equal field arguments, one explicitly obtains the
expressions of the functions $\widehat{\Gamma}_k^{(2)}$,
$\widetilde{\Gamma}_k^{(2)}$, $\Gamma_{k,abc}^{(3)}$,
$\Gamma_{k,abcd}^{(4)}$ appearing in
Eqs. (\ref{eq_unif_Gamma2}-\ref{eq_unif_Gamma4}):
\begin{equation}
  \label{eq_scaled_hatGamma2}
\widehat{\Gamma}_k^{(2)}(\phi; q=0)\simeq \frac{k^{4-\bar\eta}}{T_k} u_{*}''(\varphi),
\end{equation}
\begin{equation}
\label{eq_scaled_tildeGamma2}
\widetilde{\Gamma}_k^{(2)}(\phi,\phi; q=0)\simeq -\frac{k^{4-\bar\eta}}{T_k^2}\delta_{*,0}(\varphi),
\end{equation}
\begin{equation}
\label{eq_scaled_Gamma3}
\begin{split}
\Gamma_{k,abc}^{(3)}(\phi)\simeq \frac{k^{4-\bar\eta-\frac{1}{2}(d-4+\bar\eta)}}{T_k}\bigg\{&\delta_{abc} u_{*}'''(\varphi)  - \\&\frac {\delta_{*,0}'(\varphi)}{2T_k}(\delta_{ab}+ \delta_{bc}+ \delta_{ca}) \bigg\},
\end{split}
\end{equation}
\begin{equation}
\label{eq_scaled_Gamma4}
\begin{split}
  \Gamma_{k,abcd}^{(4)}(&\phi)\simeq \frac{k^{4-\bar\eta-(d-4+\bar\eta)}}{T_k}\bigg\{\delta_{abcd}
  u_{*}''''(\varphi) - \frac
  {\delta_{*,0}''(\varphi)}{4T_k}\big[\delta_{abc}\\&+ \delta_{bcd}+
  \delta_{cda}+ \delta_{dab}+\delta_{ab}\delta_{cd} +
  \delta_{ac}\delta_{bd}+ \delta_{ad}\delta_{bc}\big] \\& - \frac
  {f^{(02)}(\varphi,0)}{4T_k^2}\big[\delta_{abc}+ \delta_{bcd}+
  \delta_{cda}+ \delta_{dab}\\&-(\delta_{ab}\delta_{cd} +
  \delta_{ac}\delta_{bd}+ \delta_{ad}\delta_{bc}) \big]\bigg\},
\end{split}
\end{equation}
where of course the dimensionful field $\phi$ vanishes (as
$k^{(d-4+\bar\eta)/2)}$).

Inserting the above equations in the expressions of the connected
Green functions,
Eqs. (\ref{eq_truncation_W2ab}-\ref{eq_truncation_tildeW2}) and their
generalizations for higher-order functions, finally leads to
Eqs. (\ref{eq_scaled_hatG2}-\ref{eq_scaled_G4}) of the text, with for
instance
\begin{equation}
\label{eq_g2star}
g_{*}^{(2)}(\varphi)=\delta_{*,0}(\varphi)u_{*}''(\varphi)^{-2}
\end{equation}
\begin{equation}
\label{eq_h2star}
h_{*}^{(2)}(\varphi)=u_{*}''(\varphi)^{-1},
\end{equation}
etc... As stated in the text, the expressions for the other functions
$g_{*}^{(p)}(\varphi)$, $h_{*}^{(p)}(\varphi)$, $p=3,4,...$, only
involve $u_*''(\varphi)$, $\delta_{*,0}(\varphi)$ and their
derivatives. Obtaining them is tedious but straightforward, and the
resulting formulas are not worth displaying.

\end{document}